\documentclass{ptephy_v1}
\usepackage{latexsym,graphicx,amssymb,amsmath,mathrsfs}
\usepackage{multirow}
\usepackage{epstopdf}
\usepackage{url}
\graphicspath{{./Figs/}}
\usepackage{bm}
\usepackage[normalem]{ulem}

\renewcommand\sout[1]{\bgroup \color{red} \ULdepth=-.5ex \ULset {#1}}
\renewcommand{\comment}[1]{}
\graphicspath{{./Figs/}}
\newcommand{\nk}{{n\bm{k}}}
\newcommand{\bx}{{\bm{x}}}
\newcommand{\bk}{{\bm{k}}}
\newcommand{\tx}{{\tau\bx}}

\newcommand{\VEV}[1]{\left\langle{#1}\right\rangle}
\newcommand{\VEVT}[1]{\left\langle{#1}\right\rangle_T}
\newcommand{\VEVev}[1]{\left\langle{#1}\right\rangle}

\newcommand{\Zpf}{\mathcal{Z}}
\newcommand{\Hml}{\mathcal{H}}
\newcommand{\hyph}{\operatorname{-}}

\newcommand{\TL}{2{\hyph}loop}
\newcommand{\Trep}{T_\mathrm{repl}}
\begin{document}
\preprintnumber{YITP-20-94, KUNS-2825}
\title{Replica evolution of classical field in 4+1 dimensional spacetime
toward real time dynamics of quantum field~\footnote{Report number: YITP-20-94, KUNS-2825}}
\author[1]{Akira Ohnishi}
\affil{Yukawa Institute for Theoretical Physics, Kyoto University,
Kyoto 606-8502, Japan \email{ohnishi@yukawa.kyoto-u.ac.jp}}
\author[2]{Hidefumi Matsuda}
\affil{Department of Physics, Faculty of Science, Kyoto University,
Kyoto 606-8502, Japan}
\author[1]{Teiji Kunihiro}
\author[3]{Toru T. Takahashi}
\affil{National Institute of Technology, Gunma college, Gunma 371-8530, Japan}
\begin{abstract}
Real-time evolution of replicas of classical field
is proposed as an approximate simulator of real-time quantum field dynamics
at finite temperatures.
We consider $N$ classical field configurations,
$(\phi_{\tx},\pi_{\tx}) (\tau=0,1,\cdots N-1)$, dubbed as replicas,
which interact with each other via the $\tau$-derivative terms
and evolve with the classical equation of motion.
The partition function of replicas is found to be proportional
to that of quantum field in the imaginary time formalism.
Since the replica index can be regarded as the imaginary time index,
the replica evolution is technically the same as the molecular dynamics part
of the hybrid Monte-Carlo sampling.
Then the replica configurations should reproduce the correct
quantum equilibrium distribution after the long-time evolution.
At the same time, evolution of the replica-index average
of field variables is described by the classical equation of motion
when the fluctuations are small.
In order to examine the real-time propagation properties of replicas,
we first discuss replica evolution in quantum mechanics.
Statistical averages of observables are precisely obtained
by the initial condition average of replica evolution,
and the time evolution of the unequal-time correlation function,
$\langle x(t) x(t')\rangle$,
in a harmonic oscillator
is also described well by the replica evolution in the range $T/\omega > 0.5$.
Next, we examine the statistical and dynamical properties
of the $\phi^4$ theory in the 4+1 dimensional spacetime,
which contains three spatial, one replica index or the imaginary time,
and one real time.
We note that the Rayleigh-Jeans divergence can be removed in replica evolution
with $N \geq 2$ when the mass counterterm is taken into account.
We also find that the thermal mass obtained from
the unequal-time correlation function at zero momentum
grows as a function of the coupling as in the perturbative estimate
in the small coupling region.
\end{abstract}
\maketitle
\section{Introduction}

Classical dynamics has been utilized to understand
non-equilibrium evolution of quantum many-body systems in various fields of 
physics~\cite{TDGP,TDHF,inflation,CSS,CYM,Dumitru-Nara,YM-chaos,Entropy,Shear,Matsuda}.
The classical equation of motion for the phase space distribution
(Vlasov equation)~\cite{Vlasov} is known to provide an approximate solution
of the quantum equation of motion for the density matrix
(von Neumann equation)~\cite{vonNeumann},
provided that the classical analogue of the quantum mechanical
distribution function~\cite{Wigner} is given as the initial condition
and the $\mathcal{O}(\hbar^2)$ effects are negligible.
It is also known that one can numerically obtain the solution
of the Vlasov equation by solving the classical equations of motion
for particle ensemble representing the phase space distribution~\cite{TP}.
This favorable feature of classical dynamics has been invoked also in field
theories~\cite{inflation,CSS,CYM,Dumitru-Nara,YM-chaos,Entropy,Shear,Matsuda}.
For example, the classical Yang-Mills (CYM) field
has been adopted to describe the initial stage of high-energy heavy-ion
collisions, and have provided important insights into the non-equilibrium
dynamics of the gluon 
field~\cite{CYM,Dumitru-Nara,YM-chaos,Entropy,Matsuda}.

Compared with the successes in the far-from-equilibrium stages,
the applicability of classical dynamics is limited
when discussing equilibrium properties of quantum systems.
Since the equipartition law applies to classical equilibrium,
the number of high-momentum particles is overestimated
and one encounters the Rayleigh-Jeans divergence.
One possible way to manage the divergence is
treating the hard modes above the cutoff separately.
By integrating hard modes~\cite{Bodeker:1995pp,Greiner:1996dx}
or by introducing the mass counterterm~\cite{Aarts},
one can obtain the effective action of the classical field,
soft modes below the cutoff,
and utilize the action to evaluate the evolution.
In CYM theory, dynamical evolution of the coupled system
of classical field and particles is explicitly solved and was demonstrated
to promote equilibration~\cite{Dumitru-Nara}.
Still, classical field obeys classical statistics,
then the cutoff momentum should be chosen to be of the order of $T$
or smaller also in these frameworks.
While the two-particle irreducible (2PI) effective action approach
can treat classical field and particles
on the same footing~\cite{Berges:2004yj,Aarts:2001yn,Hatta-Nishiyama},
the numerical cost is large and it is not yet easy to apply
to realistic systems under inhomogeneous classical field.

Thus it is desirable to develop frameworks
which inherit the merit of classical field dynamics
but properly describe quantum statistical equilibrium
after a long-time evolution.
Including these two features is known to be important 
in nuclear transport phenomena~\cite{QStat},
and it is desirable also to describe non-equilibrium phenomena in field theories
as mentioned above.
In the stochastic quantization~\cite{SQ},
one can obtain field configurations $\{\phi\}$ by solving the Langevin equation,
$d\phi_x/dt=-\partial S/\partial \phi_x + \zeta_x$,
with $t$ being the fictitious time and $\zeta_x$ being the white noise,
$\VEV{\zeta_x(t)\zeta_y(t')}=2\delta_{xy}\delta(t-t')$.
The field distribution approaches the quantum one,
while the above Langevin equation cannot be regarded as
the equation of motion to describe the real time evolution.
There is a hint to incorporate the quantum statistical
property into the real time evolution
in the imaginary time formalism of finite temperature quantum field theory,
where the field variables in 3D space are enlarged to those in 3+1D
spacetime introducing the imaginary time. 
In the path integral representation, the thermally equilibrated
quantum field distribution
is described by $\exp(-S[\phi])$,
where $S[\phi]$ is the 3+1D Euclidean action.
In the molecular dynamics part of the hybrid Monte-Carlo (HMC)
sampling~\cite{HMC},
the Hamiltonian is set to be
$\Hml=\sum\pi_{\bm{x}\tau}^2/2+S[\phi]$
with $\pi_{\bm{x}\tau}$ being the canonical conjugate of the field variable $\phi_{\bm{x}\tau}$
at a spacetime point $(\bm{x},\tau)$,
where $\tau$ is the imaginary time coordinate.
The classical equation of motion,
$d\phi_{\bm{x}\tau}/dt=\partial\Hml/\partial\pi_{\bm{x}\tau}$
and $d\pi_{\bm{x}\tau}/dt=-\partial\Hml/\partial\phi_{\bm{x}\tau}$,
is solved with the initial condition of $\VEV{\pi_{\bm{x}\tau}^2}=1$,
where the {\em time} variable $t$ is the fictitious simulation time
and introduced in an {\em ad hoc} manner.
After a long-time evolution, the system reaches the equilibrium
described by the classical partition function at temperature of unity,
$\Zpf=\int\mathcal{D}\pi\,\mathcal{D}\phi\,\exp(-\Hml)
\propto\int\mathcal{D}\phi\,\exp(-S[\phi])$,
then we can correctly sample the quantum field configuration in equilibrium.

\begin{figure}[htbp]
\begin{center}
\includegraphics[width=8cm,bb=20 20 770 550,clip]{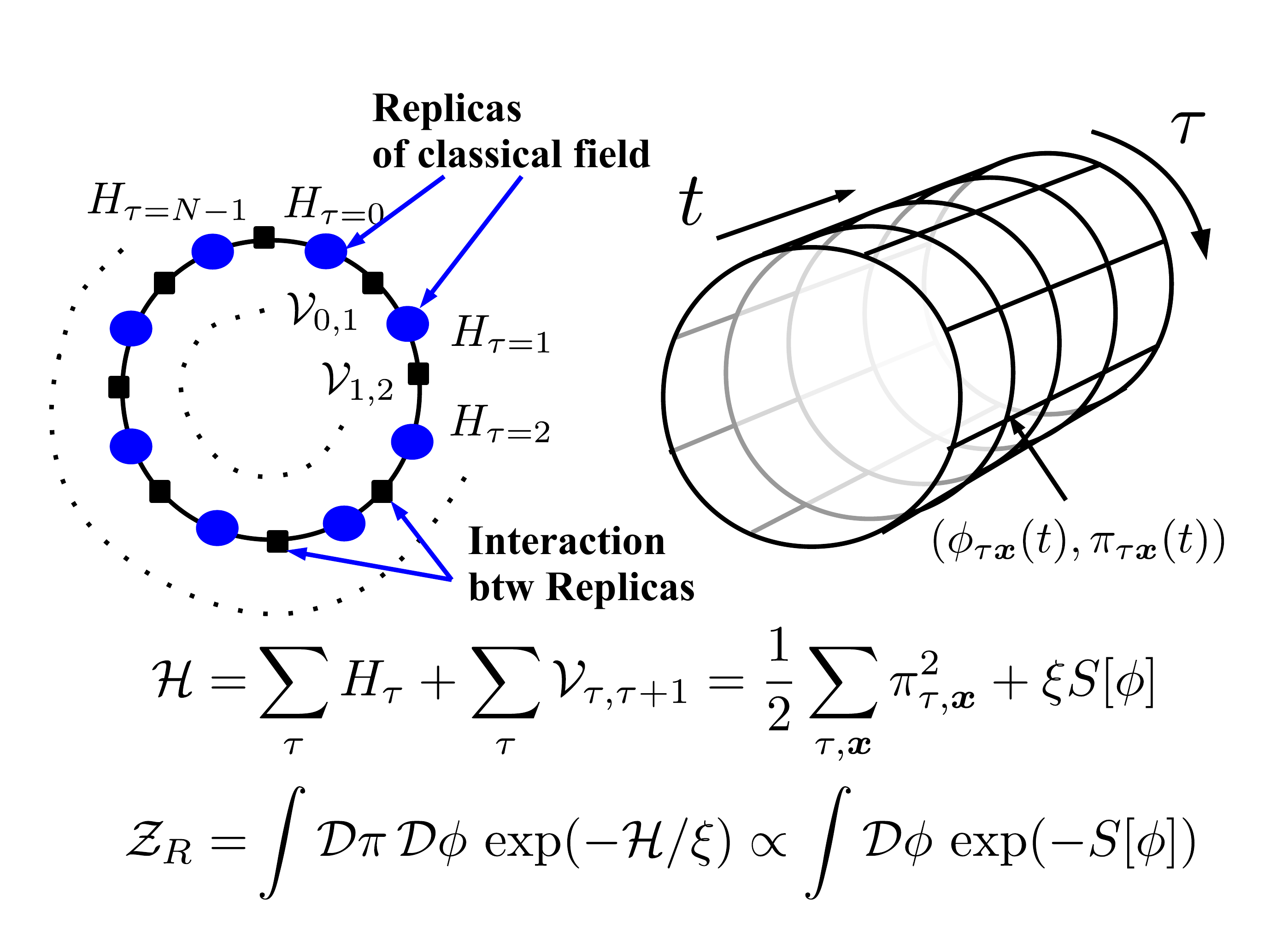}
\end{center}
\caption{Replicas and their evolution.
Replica configuration $(\phi_{\tau\bm{x}},\pi_{\tau\bm{x}})$ evolves
with the classical equation of motion using the Hamiltonian $\mathcal{H}$.
The interaction part of the replica Hamiltonian $\mathcal{H}$
is chosen so that the $\phi$ part of $\mathcal{H}$
agrees with the Euclidean action $S[\phi]$
multiplied by the lattice anisotropy $\xi=a/a_\tau$.
Thus the replica partition function $\mathcal{Z}$
becomes proportional to the equilibrium quantum field partition function,
$\int \mathcal{D}\exp(-S[\phi])$.
}
\label{Fig:Replica}
\end{figure}

In this article, we examine the time evolution of the enlarged field variables
in the 3+1 dimensions, $(\phi_{\bm{x}\tau},\pi_{\bm{x}\tau})$,
a set of 3D classical field configurations
referred to as \textit{replicas},
and propose that this can be regarded as the real time evolution
of 3D quantum field in equilibrium.
Each replica with the index $\tau$ corresponds to the field configuration
on each (discretized) imaginary time coordinate
in the finite temperature quantum field theory,
and then the replica evolution is practically 
equivalent to the molecular dynamics part of HMC.
As schematically shown in Fig.~\ref{Fig:Replica},
the replica Hamiltonian $\Hml$
is given as the sum of the Hamiltonian of each replica
plus the interactions between the 
nearest neighbor replicas, $\mathcal{V}_{\tau,\tau+1}$,
which is referred to as the $\tau$-derivative term
and causes thermalization of replicas.
The distribution of one replica configuration
regarded as the \textit{system} relaxes to the quantum equilibrium
distribution by the $\tau$-derivative interactions with
other replicas regarded as the \textit{heat bath}.
Thus, on the one hand, field variables $\phi$ reaches
correct quantum statistical distribution after a long-time replica evolution.
On the other hand, the replica-index average of field variables obeys
the standard classical field equation of motion
when fluctuations are small as shown later,
and then the variable $t$, the fictitious simulation time in HMC,
works as the real time variable in the replica evolution.
Hence the replica evolution can reproduce
both the quantum statistics and the classical evolution in these two limits.
These features of the replica evolution are encouraging for us to 
consider it as a candidate which describes dynamical evolution of quantum field,
while the formal {\em justification} of the replica evolution 
as a quantum time evolution has not been found.
In order to examine the validity of replica evolution,
we investigate the real-time evolution of
the unequal-time two-point function at zero momentum
numerically without and with the mass counterterm.
We demonstrate that the thermal mass obtained from the replica evolution
is consistent with the perturbative calculation results
in quantum field theory.
While the longer term goal is to apply the replica evolution
to the non-equilibrium real time evolution of a system of fields,
we here concentrate on the equilibrium features as the first step toward the goal.

This article is organized as follows.
In Sec.~\ref{Sec:HO},
we introduce the replica evolution in quantum mechanics,
and the statistical and dynamical properties of a harmonic oscillator
are examined.
In Sec.~\ref{Sec:Theory},
we introduce the replica evolution in classical field,
and the statistical properties of replicas are examined
in the free field case.
We also discuss the mass counterterm on the lattice.
In Sec.~\ref{Sec:Results},
we show the results of real-time evolution of replicas in the $\phi^4$ theory.
By using the damped oscillator ansatz,
we fit the unequal-time two-point function of replica evolution,
and compare the obtained thermal mass and damping rate
with the perturbative calculation results.
Section~\ref{Sec:Summary} is devoted to the summary and perspectives.

\section{Replica evolution in quantum mechanics}

In this section, we introduce the replica evolution in quantum mechanics.
We consider $N$ sets of canonical variables,
which are labeled by the replica index $\tau$ and are the functions of time $t$,
$(\bm{x}_\tau(t),\bm{p}_\tau(t))$.
As shown in Fig.~\ref{Fig:Replica}, there are two variables, $\tau$ and $t$,
which are related to time:
The variable $\tau$ is proportional to the continuous temporal variable $\bar{\tau}$,
$\tau/\xi\to\bar{\tau}$, in the large $N$ limit
in the imaginary time formalism,
and we regard it as an index of the replica.
The time $t$ corresponds to the fictitious simulation time in HMC,
and 
is found to work as the real time in the replica evolution.

\subsection{Statistics and dynamics of replicas in quantum mechanics}

In this section, we introduce replica evolution in quantum mechanics
and examine its statistical and dynamical properties.
We consider the classical system described by the following Hamiltonian
\begin{align}
H(\bm{x},\bm{p}) = & \frac{\bm{p}^2}{2} +V(\bm{x})
=\sum_{i=1}^D \frac{p_i^2}{2}+V(\bm{x})
,\label{Eq:qm}
\end{align}
where the mass is set to be unity for simplicity,
and $D$ is the number of components in $\bm{x}$ and $\bm{p}$.
We now consider $N$ replicas of canonical variables,
$(\bm{x}_\tau,\bm{p}_\tau) (\tau=0,1,\cdots, N\!-\!1)$,
whose Hamiltonian is given by the sum of the Hamiltonians
$H(\bm{x}_\tau,\bm{p}_\tau)$
over the replica indices $\tau$ and the $\tau$-derivative terms,
\begin{align}
\Hml
=&\sum_{\tau} H(\bm{x}_\tau,\bm{p}_\tau)+\mathcal{V}
\ ,\quad
\mathcal{V}
=\frac{\xi^2}{2}\sum_{\tau}(\bm{x}_{\tau+1}-\bm{x}_\tau)^2
\ .
\end{align}
The periodic boundary condition is imposed in the $\tau$ direction,
$(\bm{x}_N,\bm{p}_N)=(\bm{x}_0,\bm{p}_0)$.
The replicas are assumed to evolve in real time $t$
according to the canonical equations of motion,
\begin{align}
\frac{d\bm{x}_\tau}{dt}=&\frac{\partial\Hml}{\partial \bm{p}_\tau} 
\ ,\quad
\frac{d\bm{p}_\tau}{dt}=-\frac{\partial\Hml}{\partial \bm{x}_\tau}
\ .\label{Eq:EOM_QM}
\end{align}
The replica evolution with Eq.~\eqref{Eq:EOM_QM} has two distinct features:
First, the ensemble of replica configurations
follows the quantum statistical distribution 
of the spatial variables $x=\{\bm{x}_\tau|\tau=0,1,\cdots,N-1\}$
in the long-time evolution.
Second, the evolution of replica-index average agrees
with the purely classical evolution,
when the fluctuations among replicas are small.

Let us examine the first point on the statistics of replicas.
The partition function of replicas at temperature $T_\mathrm{repl}=\xi$
is given as
\begin{align}
\Zpf_R(\xi)
=&\int \prod_\tau\frac{d\bm{x}_\tau d\bm{p}_\tau}{(2\pi)^D} e^{-\Hml/\xi}
=\frac{\xi^{ND/2}}{(2\pi)^{ND}} \int \prod_\tau d\bm{x}_\tau e^{-S(x)}
\ ,\\
S(x)
=& \frac{1}{\xi}\sum_\tau\left[
\frac{\xi^2}{2}\left(\bm{x}_{\tau+1}-\bm{x}_\tau\right)^2 
+ V(\bm{x}_\tau)
\right]
\underset{N\to \infty}{\longrightarrow}
S_E[x]=\int_0^\beta d\bar{\tau} 
L_E\left(x,\frac{\partial x}{\partial\bar{\tau}}\right)
\ ,\\
L_E
=&
\frac12 \left[\frac{\partial \bm{x}(\bar{\tau})}{\partial \bar{\tau}}\right]^2 
+ V(\bm{x}(\bar{\tau}))
\end{align}
where $\beta=N/\xi$ and 
$\tau/\xi\to\bar{\tau}$ is the continuous imaginary time.
Since $S_E[x]$ is the Euclidean action at temperature $T=1/\beta=\xi/N$
in the imaginary time formalism, the replica partition function $\Zpf_R$
at temperature $T_\mathrm{repl}=\xi$
is proportional to the quantum mechanical partition function 
at $T=\xi/N$ in the large $N$ limit.
Thus observables as functions of $x$ in 
quantum equilibrium are correctly obtained
from the thermal average of observables
$\VEVT{\mathcal{O}(\bm{x})}$
in classical equilibrium of replica configurations,
\begin{align}
\VEV{\mathcal{O}(\bm{x})}_{T}
\equiv&\VEV{\mathcal{O}(\bm{x_{\tau'}})}
=\frac{1}{\Zpf_R(\xi)}\int \prod_\tau\frac{d\bm{x}_\tau d\bm{p}_\tau}{(2\pi)^D}
\,e^{-\Hml/\xi}\,\mathcal{O}(\bm{x}_{\tau'})
\nonumber\\
=&
\int\prod_\tau{d\bm{x}_\tau}\,e^{-S(x)}
\,\widetilde{\mathcal{O}}(\bm{x})
\,/\,
\int\prod_\tau{d\bm{x}_\tau}\,e^{-S(x)}
,\label{Eq:Obs_QMrep}
\end{align}
where $\tau'$ in the first line is the replica index of observation,
and in the second line, we replace $\mathcal{O}(\bm{x}_{\tau'})$
with the ``replica-index average'',
\begin{align}
\widetilde{\mathcal{O}}(\bm{x},\bm{p})
= \frac1N\sum_\tau\mathcal{O}(\bm{x}_\tau,\bm{p}_\tau),
\end{align}
using the translational invariance in the $\tau$-direction.
Then, a thermal expectation value of an operator $\VEVT{O(\bm{x})}$
 can be obtained as an expectation value of the replica-index averaged
 operator,
\begin{align}
\VEVT{\mathcal{O}(\bm{x})} = \VEV{\widetilde{\mathcal{O}}(\bm{x})}.
\end{align}
 
In the imaginary time formalism, the thermal average of an observable
in quantum mechanics is given as
\begin{align}
\VEV{\mathcal{O}(\bm{x})}=&\mathrm{tr}(\mathcal{O}(\bm{x})\,e^{-\beta H})/\mathrm{tr}(e^{-\beta H})
=\int \mathcal{D}x e^{-S_E[x]}\,\mathcal{O}(\bm{x}(\bar{\tau}'))
/ \int \mathcal{D}x e^{-S_E[x]}
,\label{Eq:Obs_QM}
\end{align}
where the imaginary time of observation $\bar{\tau}'$ appears
in the path integral representation.
After replacing $\mathcal{O}(\bm{x}(\bar{\tau}'))$
with its imaginary time average,
the quantum statistical average \eqref{Eq:Obs_QM}
is found to be described by the replica average \eqref{Eq:Obs_QMrep}
in the large $N$ limit.

The classical equilibrium of replica configurations
can be generally obtained by the long-time evolution
with the canonical equation of motion, Eq.~\eqref{Eq:EOM_QM},
due to the chaoticity of the system. 
Practically, it is useful to take the ``replica ensemble average''
instead of the long-time average, since there is no autocorrelation
in the former. We prepare $N_\mathrm{conf} (\gg 1)$
initial replica configurations
$\left\{(x^{(i)},p^{(i)})|i=1,2,\cdots, N_\mathrm{conf}\right\}$ at $t=0$
appropriately,
solve the equation of motion, then the replica average is calculated as
\begin{align}
\VEV{\mathcal{O}(\bm{x})} 
\simeq& \lim_{t\to\infty}\frac{1}{N_\mathrm{conf}}\sum_{i=1}^{N_\mathrm{conf}}
\widetilde{\mathcal{O}}(x^{(i)}(t))
=\lim_{t\to\infty}
\frac{1}{N_\mathrm{conf}}\sum_{i=1}^{N_\mathrm{conf}}
\frac{1}{N}\sum_{\tau=0}^{N-1}
\mathcal{O}(\bm{x}_\tau^{(i)}(t))
.
\end{align}

Let us turn to the second point.
While replica ensemble simulates quantum statistical ensemble
after a long-time evolution,
the replica-index averages of the canonical variables evolve
as the classical variables
when the fluctuations among replicas are small.
The equation of motion for the replica-index average
of the canonical variables reads 
\begin{align}
\frac{d\widetilde{\bm{x}}}{dt}
=&\frac{1}{N}\sum_\tau \frac{d\bm{x}_\tau}{dt}
=\frac{1}{N}\sum_\tau \bm{p}_\tau=\widetilde{\bm{p}}
\label{Eq:EOM_tildex}
,\\
\frac{d\widetilde{\bm{p}}}{dt}
=&\frac{1}{N}\sum_\tau \frac{d\bm{p}_\tau}{dt}
=-\frac{1}{N}\sum_\tau \frac{\partial \Hml}{\partial \bm{x}_\tau}
=-\frac{1}{N}\sum_\tau \frac{\partial V(\bm{x}_\tau)}{\partial \bm{x}_\tau}
\nonumber\\
=&-\frac{\partial V(\widetilde{\bm{x}})}{\partial \widetilde{\bm{x}}}
- \frac{1}{2}\sum_{i,j} \frac{\partial^3 V(\widetilde{\bm{x}})}
{\partial \widetilde{\bm{x}}\partial\widetilde{x}_i\partial\widetilde{x}_j}
\frac{1}{N}\sum_\tau(x_{i\tau}-\widetilde{x}_i)(x_{j\tau}-\widetilde{x}_j)
+\mathcal{O}((\delta x)^3)
\nonumber\\
=&-\frac{\partial V(\widetilde{\bm{x}})}{\partial \widetilde{\bm{x}}}
+\mathcal{O}((\delta x)^2)
,\label{Eq:EOM_tildep}
\end{align}
where $x_{i\tau}$ is the $i$-th component of $\bm{x}_\tau$
and $(\delta x)^2=\sum_\tau(\bm{x}_\tau-\widetilde{\bm{x}})^2/N$
is the fluctuations of $\bm{x}_\tau$ in one configuration of replicas.
It should be noted that the $\tau$-derivative terms do not operate
because of the periodic boundary condition,
and the second derivative terms of $V$ disappear from the definition
of $\widetilde{\bm{x}}$.
It is interesting to find that the first line of Eq.~\eqref{Eq:EOM_tildep}
shows the Ehrenfest's theorem
\begin{align}
\frac{d^2\VEV{x}}{dt^2}=-\VEV{\frac{\partial V(x)}{\partial x}},\label{Eq:quan}
\end{align}
where $\VEV{\cdots}$ denotes the replica-index average here.
Then when $(\delta x)^2$ is small,
these equations of motion tell us the classical
nature of the replica-index average,
\begin{align}
\frac{d^2\widetilde{\bm{x}}}{dt^2}
\simeq& -\frac{\partial V(\widetilde{\bm{x}})}{\partial \widetilde{\bm{x}}}
.
\end{align}
By comparison, $(\delta x)^2$ should be a part of quantum fluctuations.

\subsection{Replica evolution of harmonic oscillator}
\label{Sec:HO}

We now look further into the dynamical property of replicas
of a single harmonic oscillator.
By choosing $V=\omega^2 x^2/2$ and $D=1$ in Eq.~\eqref{Eq:qm},
the Hamiltonian is given as
\begin{align}
H(x,p) = & \frac12 p^2 + \frac{\omega^2}{2}x^2
=\omega\left(a^\dagger a+\frac12\right)
,\quad
a=\frac{1}{\sqrt{2}}\left(\sqrt{\omega}x + \frac{ip}{\sqrt{\omega}}\right)
.
\end{align}

In order to investigate the statistical properties of replicas,
it is useful to adopt the Fourier transform with respect
to the replica index,
\begin{align}
\bar{x}_n=\frac{1}{\sqrt{N}}\sum_\tau e^{i\omega_n\tau} x_\tau
\ ,\quad
\bar{p}_n=\frac{1}{\sqrt{N}}\sum_\tau e^{i\omega_n\tau} p_\tau
\ ,
\end{align}
where $\omega_n=2\pi n/N$ denotes the Matsubara frequency.
With $(\bar{x}_n,\bar{p}_n)$,
the Hamiltonian is represented by that of the $N$ free harmonic oscillators,
\begin{align}
\Hml
=&\sum_{n}\left[\frac12 \bar{p}_n^2 + \frac{M_n^2}{2} \bar{x}_n^2\right]
\ ,\quad
M_n^2=\,\omega^2 + 4\xi^2\sin^2(\omega_n/2)
\ .
\end{align}
The replica partition function is obtained as
\begin{align}
\Zpf_R(\xi)
=&\prod_n \left(
\int\frac{d\bar{x}_n d\bar{p}_n}{2\pi}
e^{-\bar{p}_n^2/2\xi- M_n^2\bar{x}_n^2/2\xi}
\right)
=\prod_n \left(
\frac{\xi}{M_n}
\right)
\ .
\label{Eq:ZfreeHOR}
\end{align}
By using the Matsubara frequency summation formula
explained in Appendix~\ref{Sec:Matsubara},
the logarithm of the partition function is found to be
\begin{align}
-\log\Zpf_R(\xi)
=&\sum_n \log(M_n/\xi)
=\frac12\sum_n \log\left[
\left(\frac{\omega}{2\xi}\right)^2+\sin^2(\omega_n/2)
\right]
+N \log 2
\nonumber\\
=&\log\left[2\sinh\left(\frac{\Omega}{2T}\right)\right]
\underset{N\to \infty}{\longrightarrow}
\log\left[2\sinh\left(\frac{\omega}{2T}\right)\right]
\ ,
\label{Eq:ZfreeHORlog}
\end{align}
where $\Omega$ is given as
\begin{align}
\Omega
&=2\xi\,\mathrm{arcsinh}\,(\omega/2\xi)
=2NT\,\mathrm{arcsinh}\,(\omega/2NT)
\underset{N\to \infty}{\longrightarrow}
\omega
\ .
\end{align}
The replica partition function, Eq.~\eqref{Eq:ZfreeHOR}, at large $N$
agrees with the quantum mechanical partition function at $T=\xi/N$,
\begin{align}
\Zpf_Q(T)
=&\sum_{n=0}^\infty e^{-E_n/T}
=\sum_{n=0}^\infty e^{-\omega/T(n+1/2)}
=\frac{e^{-\omega/2T}}{1-e^{-\omega/T}}
=\left[2\sinh\left(\frac{\omega}{2T}\right)\right]^{-1}
\ .
\end{align}
Thus we find that it is possible to obtain quantum statistical results
by using a replica ensemble in equilibrium,
a large number of configurations of $N$ sets of canonical variables,
$(x_\tau,p_\tau)$, which are prepared to be in classical equilibrium.

\begin{figure}[htbp]
\begin{center}
\includegraphics[width=10cm,bb=0 0 360 250,clip]{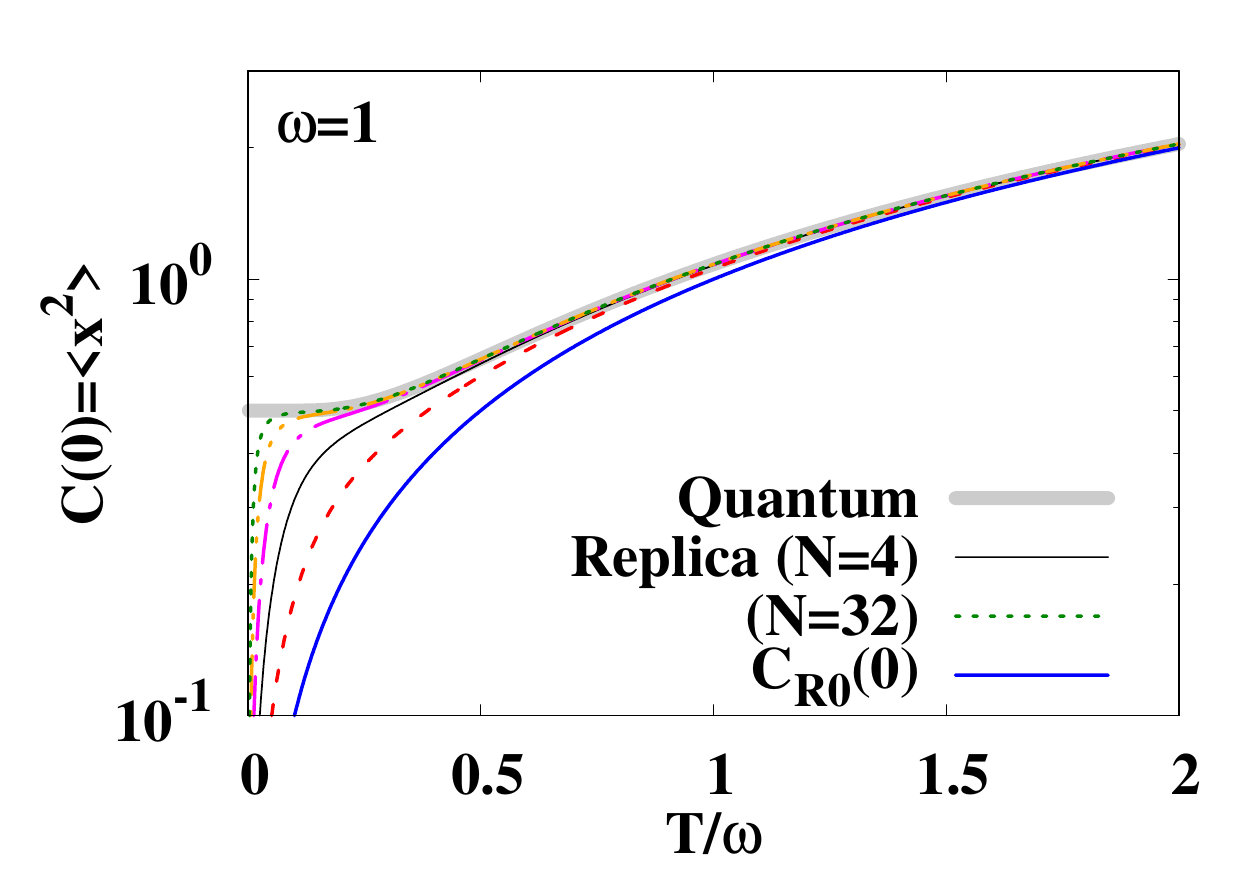}
\end{center}
\caption{Equal-time correlation function of a harmonic oscillator.
Grey solid curve shows the quantum mechanical results ($C_Q$),
the blue solid curve shows the classical field results ($C_{R0}$),
and other curves show the replica evolution results ($C_R$, $N >1$);
red dashed, black solid, magenta dot-dashed, orange dot-dot-dashed,
and green dotted curves show the results with
$N=2, 4, 8, 16$ and 32, respectively.
}
\label{Fig:HO-xsq}
\end{figure}

We next examine the time-evolution of replicas
by using the time-correlation function, $C(t)=\VEV{x(t)x(0)}$.
For preparation, let us recall the quantum mechanical results.
The time correlation of the spatial coordinate is calculated as
\begin{align}
&\VEV{\,\psi|x_H(t)x_H(0)|\psi\,}
=\VEV{\,\psi|e^{iHt}xe^{-iHt}x|\psi\,}
\nonumber\\
=&\frac{1}{2\omega}\sum_{n,n'} c_n^* c_{n'}
\VEV{n|e^{iHt}(a+a^\dagger)e^{-iHt}(a+a^\dagger)|n'}
\nonumber\\
=&\frac{1}{2\omega}\sum_{n} 
\left\{
c_n^* c_{n}
\left[ne^{i\omega t}+(n+1)e^{-i\omega t}\right]
+\sqrt{n(n-1)}
\left[c_{n-2}^* c_{n}e^{-i\omega t}+c_n^* c_{n-2}e^{i\omega t}\right]
\right\}
\ ,
\end{align}
where $x_H$ is the operator in the Heisenberg picture,
$|n\rangle$ is the energy eigen state,
$c_n$ is the expansion coefficient,
$|\psi\rangle=\sum_n c_n |n\rangle$,
and we have used the relations,
$\VEV{n\!-\!1|a|n}=\sqrt{n}$
and $\VEV{n\!+\!1|a^\dagger|n}=\sqrt{n\!+\!1}$.
In thermal equilibrium at temperature $T$,
the density matrix becomes diagonal and the statistical weights
are given by the Boltzmann factor,
\begin{align}
\rho_{nn'}=\VEVT{c_n c_{n'}^*}=\frac{e^{-\omega(n+1/2)/T}}{\Zpf_Q}\delta_{nn'}
\ .
\end{align}
Thus the time-correlation function in thermal equilibrium is obtained as
\begin{align}
C_Q(t)=&\VEVT{x_H(t)x_H(0)}
=\frac{1}{2\omega}\sum_n \frac{e^{-n\omega/T}}{\Zpf_Q}
\left[ne^{i\omega t}+(n+1)e^{-i\omega t}\right]
\nonumber\\
=&\frac{1}{2\omega}\left[\coth\left(\frac{\omega}{2T}\right)\cos\omega t-i\sin\omega t\right]
\ ,
\end{align}
where we have used the relation
$\sum_n ne^{-nx}=d/dx(\sum_n e^{-nx})$
to obtain the second line.
Since the expectation value of symmetrized product (Weyl ordering)
is obtained in classical dynamics,
we are interested in the time-even part part of the time-correlation function
given as
\begin{align}
C_Q^\mathrm{even}(t)
=&\VEVT{\frac12\,\left\{x_H(t),x_H(0)\right\}}
=\frac{\coth(\omega/2T)}{2\omega}\,\cos\omega t
\to \frac{T}{\omega^2}\,\cos\omega t
\ (T/\omega \gg 1)
\ .
\end{align}
In replica evolution,
it is easier to solve the canonical equation of motion
in the Fourier transform.
The equations of motion are,
$d\bar{x}_n/dt=\partial\Hml/\partial \bar{p}_n=\bar{p}_n$
and 
$d\bar{p}_n/dt=-\partial\Hml/\partial \bar{x}_n=-M_n^2\bar{x}_n$,
and the solution is obtained as
\begin{align}
\bar{x}_n(t)
  =\bar{x}_n(0)\cos M_nt+\frac{\bar{p}_n(0)}{M_n}\sin M_nt
\ ,\quad
\frac{\bar{p}_n(t)}{M_n}
  =-\bar{x}_n(0)\sin M_nt+\frac{\bar{p}_n(0)}{M_n}\cos M_nt
\ .
\end{align}
We evaluate thermal average imposing thermal distribution
to the initial field variables, $\bar x_n(0)$ and $\bar p_n(0)$.
Since the Boltzmann weight is given as 
$\exp(-\Hml/\xi)=\exp[-\sum_n (\bar{p}_n^2/2\xi+\bar{x}_n^2/2M_n^2\xi)]$,
the distributions of $\bar{x}_n$ and $\bar{p}_n$ in equilibrium are gaussians,
$\VEVT{\bar{x}_n(0)\bar{x}_{n'}(0)}=\xi/M_n^2 \delta_{nn'}$
and 
$\VEVT{\bar{p}_n(0)\bar{p}_{n'}(0)}=\xi \delta_{nn'}$.
Then the time-correlation function is obtained as
\begin{align}
C_R(t)=&\VEVT{x(t)x(0)}\equiv\frac{1}{N}\sum_\tau \VEVT{x_\tau(t)x_\tau(0)}
=\frac{1}{N}\sum_n \VEVT{\bar{x}_n(t)\bar{x}_n(0)}
=\frac{1}{N}\sum_n \frac{\xi}{M_n^2}\,\cos M_nt
\nonumber\\
=&\sum_n \frac{T}{M_n^2}\,\cos M_nt
\ .
\end{align}
We here adopt the \textit{replica-index average},
the average of the time-correlation function over the replica indices $\tau$,
and the \textit{replica ensemble average}, 
the average over the initial configurations.
The zero Matsubara frequency contribution in the replica formalism
agrees with the quantum mechanical result in the high-temperature limit,
\begin{align}
C_{R0}(t)
=\frac{1}{N}\VEV{\bar{x}_0(t)\bar{x}_0(0)}
=\VEV{\tilde{x}(t)\tilde{x}(0)}
=\frac{T}{\omega^2}\,\cos \omega t
\ ,
\end{align}
where $\tilde{x}=\sum_\tau x_\tau/N$ is the replica-index average of $x_\tau$.
The other Matsubara frequencies lead to higher harmonics,
which do not appear in the quantum mechanical result.
Yet they play an essential role to explain the value of equal-time correlation
$C_R(0)=\VEV{x^2(0)}$ at lower temperatures,
\begin{align}
C_R(0)=&\sum_n \frac{T}{M_n^2}
=\sum_n \frac{T}{\omega^2 + 4\xi^2\sin^2(\omega_n/2)}
=\frac{1}{2\omega}\frac{\coth(\Omega/2T)}{\sqrt{1+(\omega/2\xi)^2}}
\underset{N\to \infty}{\longrightarrow}
\frac{\coth(\omega/2T)}{2\omega}
\ .
\end{align}
In Fig.~\ref{Fig:HO-xsq}, we show the temperature dependence
of the equal-time correlation function, $C(0)=\VEV{x^2}$,
which provides the amplitude of $C(t)$.
As already mentioned, $C_{R0}(0)$ agrees with the quantum mechanical
result at high temperatures, $T \gtrsim \omega$.
At lower temperatures, 
$C_{R0}(0)$ itself deviates from the quantum mechanical result,
while the other Matsubara frequency contribution lifts up the amplitude.
As a result, the amplitude in the replica method
increases with increasing $N$, and converges to quantum mechanical result
in the large $N$ limit.
We confirm that the replica method reproduces the quantum mechanical result
of the equal-time correlation $\VEV{x^2}=C_Q(0)$ in the large $N$ limit,
which is not a total surprise since the replica method gives
correct thermal quantum distribution after a long-time evolution.

\begin{figure}[htbp]
\begin{center}
\includegraphics[width=7.5cm,bb=0 0 360 250,clip]{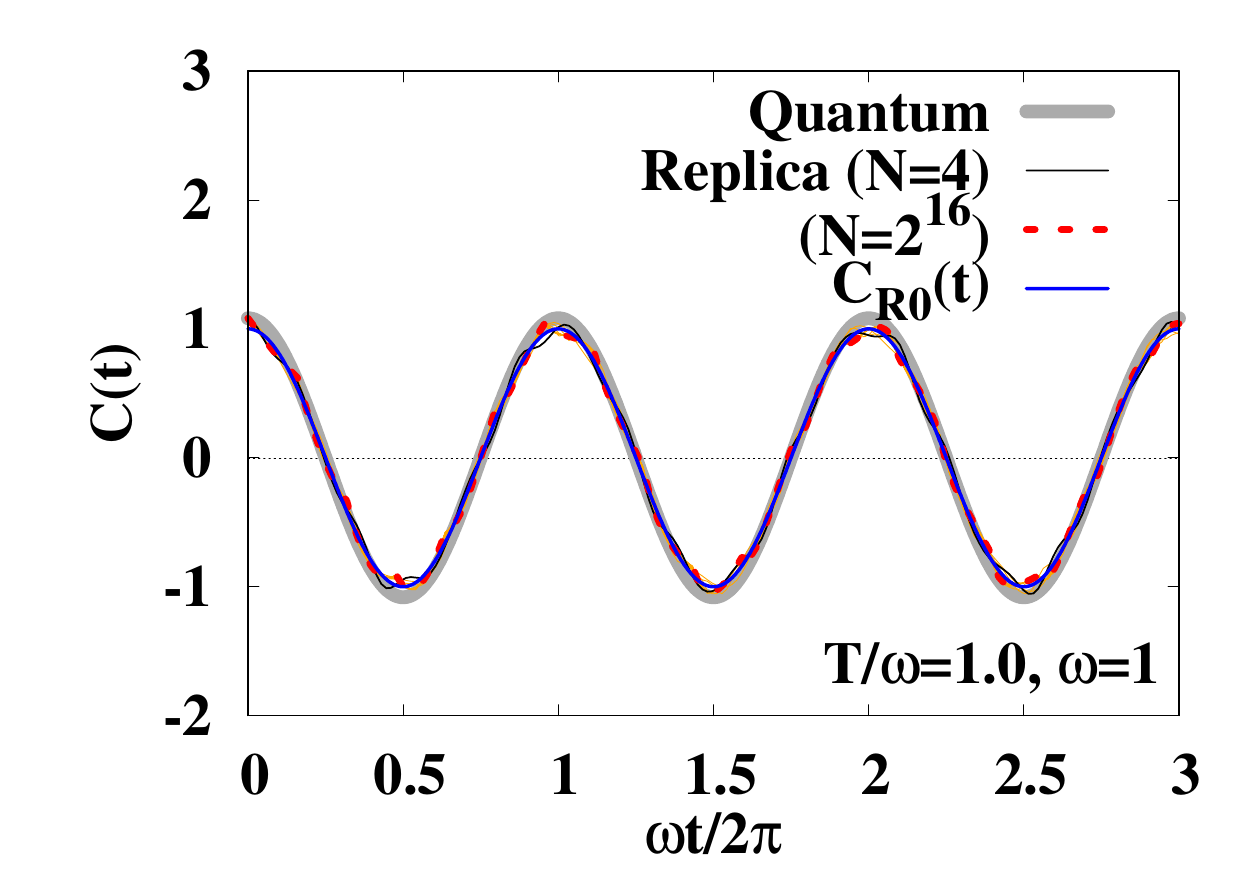}%
\includegraphics[width=7.5cm,bb=0 0 360 250,clip]{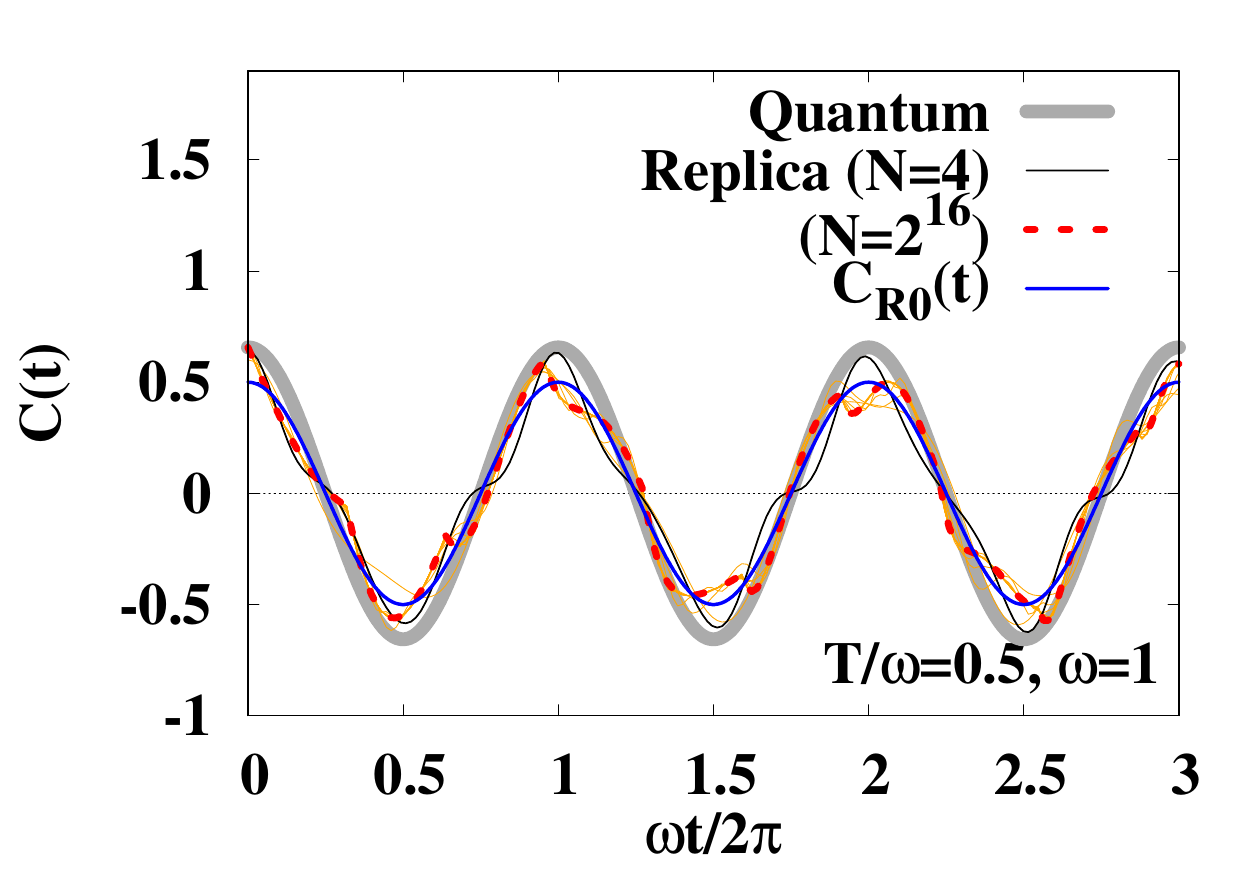}
\end{center}
\caption{Time-correlation function of a harmonic oscillator.
Grey solid curve shows the quantum mechanical results ($C_Q(t)$),
the blue solid curve shows the classical field results ($C_{R0}(t)$),
and other curves show the replica evolution results ($C_R(t)$, $N >1$);
black solid and red dotted curves show the results with $N=4$ and $N=2^{16}$,
respectively, and orange curves show other results with $N=2^{1-15}$.
}
\label{Fig:HO-T}
\end{figure}

Let us come back to the evaluation of the time-correlation functions.
In Fig.~\ref{Fig:HO-T}, we show the time-correlation function
of a harmonic oscillator in quantum mechanics
and in the replica evolution.
When the temperature $T$ is comparable to or higher than
the intrinsic frequency $\omega$,
the quantum mechanical time-correlation function is well described
by the replica evolution
which is dominated by 
the zero Matsubara frequency contribution $C_{R0}(t)$.
When $T$ is smaller than $\omega$,
the amplitude of $C_{R0}(t)$ is smaller than the quantum mechanical result
and the contributions of higher harmonics are not negligible.
As a result, $C_R(t)$ 
fluctuates around $C_{R0}(t)$ at lower temperature
as shown in the right panel of Fig.~\ref{Fig:HO-T} for $T/\omega=0.5$:
the result in the replica method with $N=4$ 
at $t=0$ agrees with that in the quantum mechanical result,
while we find deviations at $t>0$.
The deviation from the quantum result can be evaluated
by the fraction of sum of non-zero $n$ amplitudes
and the full amplitude,
\begin{align}
\frac{1}{C_R(0)}\,\sum_{n\not=0}\frac{T}{M_n^2}
=\frac{C_R(0)-C_{R0}(0)}{C_R(0)}
\underset{N\to \infty}{\longrightarrow}
\frac{C_Q(0)-C_{R0}(0)}{C_Q(0)}
=1-\frac{2T}{\omega}\tanh\left(\frac{\omega}{2T}\right)
\ ,\label{Eq:deviation}
\end{align}
which amounts to be $23.8\,\%$ ($7.6 \%$) at $T/\omega=0.5$ ($T/\omega=1$)
at large $N$.
The time-correlation function fluctuate around $C_{R0}(t)$
with maximal deviation given in Eq.\eqref{Eq:deviation}
even at very large $N$.
Yet the global behavior of the time-correlation function 
is described well by 
$C_R(t)$
in the replica formalism in the region 
$T/\omega \gtrsim 0.5$.

In this section,
we have demonstrated that the replica evolution provides configurations of $x$
in accordance with the quantum statistical distribution,
provided that the system is thermalized by some interactions with the heat bath
or the system is chaotic.
This should be also valid with any interaction,
since Eq.~\eqref{Eq:ZfreeHOR} holds for any Hamiltonian.
As for the time evolution,
the quantum mechanical time-correlation function is well described
by the replica evolution in the temperature region of $T/\omega \gtrsim 1$
and is reasonably described at $T/\omega \gtrsim 0.5$,
while the higher harmonics (non-zero $n$) distorts the time-correlation function
at $T/\omega \lesssim 1$.
It should be noted that the $n=0$ contribution is the same as
the standard classical dynamics in the harmonic oscillator,
but differences of $x$ in different replica index modify the equation of motion
even for the $n=0$ modes when interactions are switched on among replicas.

\section{Replica evolution in scalar field field}
\label{Sec:Theory}

In this section, we apply the replica evolution method
to the scalar $\phi^4$ field theory on the lattice.
Since the quantum field theory is the quantum mechanics
of field variables on spatial points, we can utilize the method
introduced in quantum mechanics also in field theories.
Thus the following discussions proceeds in parallel
to those in quantum mechanics.

\subsection{Classical Scalar Field Theory on Lattice}
We consider the $\phi^4$ theory, where the Lagrangian is given as
\begin{align}
\mathcal{L}=\frac12 \partial_\mu\phi\partial^\mu\phi
- \frac12 m^2 \phi^2 - \frac{\lambda}{24} \phi^4
\ .
\end{align}
On a $L^3$ lattice, the Hamiltonian is given as
\begin{align}
H(\phi,\pi)
=&
\sum_\bx \left[
\frac{1}{2}\pi^2_\bx
+\frac12 \left(\boldsymbol{\nabla}\phi_\bx\right)^2+
\frac{m^2}{2} \phi_\bx^2 
+ \frac{\lambda}{24}\phi^4_\bx \right]
\ ,
\end{align}
where $(\phi_\bx, \pi_\bx)$ are the canonical variables
and $\bx=(x_1,x_2,x_3)~(x_i=0, 1, \ldots, L-1)$ represents 
the lattice spatial coordinate.
Throughout this article,
we take all quantities normalized by the lattice spacing $a$.

The classical evolution of field variables is described
by the canonical equations of motion,
\begin{align}
\frac{d\phi_\bx}{dt} = \frac{\partial H}{\partial \pi_\bx}\ ,\quad
\frac{d\pi_\bx}{dt} = -\frac{\partial H}{\partial \phi_\bx}\ .
\label{Eq:EOMcl}
\end{align}
After long-time evolution, classical field distribution relaxes
to the classical statistical equilibrium,
where the classical partition function is given as,
\begin{align}
\Zpf_{cl}=\int \mathcal{D}\pi \mathcal{D}\phi\, e^{-H/T}\ .
\label{Eq:Zcl}
\end{align}
Since thermally equilibrated classical field obeys the the Rayleigh-Jeans law,
each momentum mode approximately carry the energy of $T$,
and the energy density is divergent in the continuum limit, $a^{-1}\to\infty$.
Since high-momentum modes are not suppressed by the exponential
(Boltzmann or Bose-Einstein) factor, results are sensitive to the cutoff.
Thus it is necessary to choose the cutoff appropriately
in order to deduce the results in quantum systems~\cite{Entropy,Shear,Matsuda}.
It is also possible to adopt the Hamiltonian with the mass counterterm
to avoid the divergence of the mass~\cite{Aarts},
but we cannot avoid the classical field to relax to the classical statistical 
equilibrium as long as one classical field configuration
evolves with the classical equation of motion.

\subsection{Replica evolution}

We next consider $N$ replicas of classical field,
which interact with the nearest neighbor replicas via $\mathcal{V}$,
given in the $\tau$-derivative form,
\begin{align}
&\Hml
=\sum_\tau H(\phi_\tau,\pi_\tau)
+\mathcal{V}
\ ,
\label{Eq:ReplicaH}\\
&\mathcal{V}
=\sum_\tau \mathcal{V}(\phi_\tau,\phi_{\tau+1})
=\frac{\xi^2}{2}\sum_{\tau,\bx} (\phi_{\tau+1,\bx}-\phi_{\tx})^2
,\label{Eq:ReplicaV}
\end{align}
where $(\phi_\tau,\pi_\tau)
=\left\{(\phi_{\tx},\pi_{\tx}) \mid x_i =0, 1, \ldots, L-1\right\}$
represents $\tau$-th replica of classical field
and the replica index takes the value $\tau=0,1,\cdots,N-1$.
We impose the periodic boundary condition 
in the $\tau$ direction,
$(\phi_{N\bx},\pi_{N\bx})=(\phi_{0\bx},\pi_{0\bx})$.
Provided that we solve the classical equation of motion
with the Hamiltonian $\Hml$,
\begin{align}
\dot{\phi}_{\tx} = \frac{\partial \Hml}{\partial \pi_{\tx}}\ ,\quad
\dot{\pi}_{\tx} = -\frac{\partial \Hml}{\partial \phi_{\tx}}\ ,
\label{Eq:ReplicaEOM}
\end{align}
the partition function of replicas at temperature $\Trep$
is given as
\begin{align}
\Zpf_{R}(\Trep)
=&\int \mathcal{D}\pi \mathcal{D}\phi\, e^{-\Hml/\Trep}
=\int \mathcal{D}\pi e^{-\sum_{\tx}\pi_{\tx}^2/2\Trep}
\,\int\mathcal{D}\phi\, e^{-\xi S[\phi]/\Trep}
\ ,
\label{Eq:ReplicaZTcl}\\
S[\phi]
=& \frac{1}{\xi}\sum_{\tau,\bx}\left[
\frac{\xi^2}{2}(\partial_\tau\phi_{\tx})^2
+\frac12 (\boldsymbol{\nabla}\phi_{\tx})^2
+\frac12 m^2 \phi_{\tx}^2
+\frac{\lambda}{24}\phi_{\tx}^4
\right]
,\label{Eq:ReplicaS}
\end{align}
where $\partial_\tau\phi_\tx=\phi_{\tau+1,\bx}-\phi_\tx$
and $\nabla_i\phi_\tx=\phi_{\tau,\bx+\hat{i}}-\phi_\tx$
represent the forward derivatives.
Especially, in the case where the replica temperature is chosen to be 
$\Trep=\xi$, we find
\begin{align}
\Zpf_{R}(\Trep=\xi)
=&\int \mathcal{D}\pi e^{-\sum_{\tx}\pi_{\tx}^2/2\xi}
\,\int\mathcal{D}\phi\, e^{-S[\phi]}
\ .
\label{Eq:ReplicaZ}
\end{align}
In this case, we can regard $S[\phi]$ as the Euclidean action of quantum field
in the imaginary time formalism,
where the lattice spacing in the imaginary time direction is given as
$a_\tau=a/\xi$
and the parameter $\xi$ is now interpreted as the lattice anisotropy.
The prefactor ($1/\xi$) in $S[\phi]$ shows the spacetime volume of one cell
in the lattice unit, $a^3a_\tau/a^4=1/\xi$.
The first term in the square bracket in Eq.~\eqref{Eq:ReplicaS}
can be regarded as the squared derivative of $\phi$
with respect to the continuous imaginary time given as
$[\partial\phi/\partial \bar{\tau}]^2$ with $\bar{\tau}=\tau/\xi$.
The period of the imaginary time is $N/\xi=1/T$,
where $T$ is the temperature of quantum field.
Now we find that the replica evolution of classical field
at replica temperature $\Trep=\xi=N T$
gives equilibrium quantum field configurations at temperature $T$,
\begin{align}
\Zpf_{R}(\xi)=\mathcal{N}(2\pi\xi)^{NL^3/2} Z_\mathrm{Q}(T)\, ,\ 
\Zpf_{Q}(T)=\int \mathcal{D}\phi\, e^{-S[\phi]}\, ,
\end{align}
where $\mathcal{N}$ is a normalization constant.

The difference of the present replica evolution
and the standard classical field evolution
comes from the $\tau$-derivative interaction term $\mathcal{V}$.
The equation of motion for $\phi$ in the replica evolution reads,
\begin{align}
\frac{d^2\phi_{\tx}}{dt^2}
=&-\frac{\partial H_\tau}{\partial \phi_{\tx}}
-\frac{\partial \mathcal{V}}{\partial \phi_{\tx}}
=-\frac{\partial H_\tau}{\partial \phi_{\tx}}
+\xi^2(\phi_{\tau+1,\bx}+\phi_{\tau-1,\bx}-2\phi_{\tx})
\, ,
\end{align}
where $H_\tau=H(\phi_\tau,\pi_\tau)$ and we have erased $\pi_{\tx}$
using the equation of motion.
The second term from $\mathcal{V}$, which is 
characteristic of the replica formalism,
tends to reduce the difference
of the nearest neighbor replica field variables at the same spatial point,
$\phi_{\tx}$ and $\phi_{\tau\pm1,\bx}$,
and keeps the replica ensemble in quantum equilibrium
as found in the partition function $\Zpf_R$ in Eq.~\eqref{Eq:ReplicaZ}.
Without $\mathcal{V}$,
small difference of $(\phi,\pi)$ between replicas in the initial condition
leads to very different field configurations after a long-time evolution,
since interacting classical field is generally chaotic.
As a result, classical field configurations relax to classical statistical
equilibrium and quantum thermal equilibrium cannot be kept.

As in the cases in quantum mechanics,
replica ensemble gives quantum statistical ensemble after a long-time evolution,
and the replica-index average of the classical field $\phi_{\tau \bm{x}}$
evolves like classical field.
The equation of motion for 
the replica-index average of field variables reads
\begin{align}
\frac{d\widetilde{\phi}_\bx}{dt}
=&\frac{1}{N}\sum_\tau \frac{\partial \Hml}{\partial \pi_{\tx}}
=\widetilde{\pi}_\bx
\ ,\\
\frac{d\widetilde{\pi}_\bx}{dt}
=&-\frac{1}{N}\sum_\tau \frac{\partial \Hml}{\partial \phi_{\tx}}
=\left( \boldsymbol{\nabla}^2-m^2 \right) \widetilde{\phi}_\bx
-\frac{\lambda}{3!}(\widetilde{\phi}_\bx)^3
+\mathcal{O}((\delta\phi_\bx)^2)
\, ,
\end{align}
where $(\delta\phi_\bx)^2=\sum_\tau(\phi_\tx-\widetilde{\phi}_\bx)^2/N$
denotes the variance of $\phi$ at $\bx$ in one replica configuration.
Then we get the equation of motion for $\widetilde{\phi}_\bx$,
\begin{align}
\left( \partial^2+m^2 \right) \widetilde{\phi}_\bx
\simeq
-\frac{\lambda}{3!}(\widetilde{\phi}_\bx)^3\ ,\label{Eq:eom}
\end{align}
when the fluctuations of $\phi_\bx$ in the replica configuration is small.
The equation of motion given as Eq.~(\ref{Eq:eom}) is the same as 
that for the classical field,
expectation value of $\phi(x)$.
When the fluctuations are not negligible, they modify the equation of
motion for the classical field $\widetilde{\phi}$.
For example, $(\delta\phi_\bx)^2$ contributes to the mass
as $m^2 \to m^2+\lambda(\delta\phi_\bx)^2/2$.

As in the quantum mechanics case,
the thermal average $\VEVT{\mathcal O(\phi_\bx)}$
of an arbitrary observable ${\mathcal O(\phi_\bx)}$
is defined 
as an average over the replica index $\tau$ and the thermal replica ensemble,
\begin{align}
\VEVT{\mathcal O(\phi_\bx)}
\equiv
\VEV{\widetilde{\mathcal O}(\phi_\bx)}
=&\frac{1}{\Zpf_R(\xi)}\int \mathcal{D}\pi \mathcal{D}\phi\,
  \widetilde{\mathcal{O}}(\phi_\bx)e^{-\Hml/\xi}
=\frac{1}{\Zpf_Q(T)}
 \int \mathcal{D}\phi\, \widetilde{\mathcal{O}}(\phi_\bx)e^{-S[\phi]}.
\label{Eq:ReplicaAve}
\end{align}
Since the ``classical field'' variables are obtained
by the replica-index average, fluctuations among the field configurations
with different replica indices in one replica configuration
should be regarded as a part of quantum fluctuations.
Fluctuations in replica configurations may contain
statistical and quantum fluctuations.
One replica configuration would not be enough
to describe a quantum state,
and we need at least several replica configurations
to satisfy the uncertainty principle.
Further fluctuations would be considered as statistical.
Thus taking both of the replica index and ensemble averages
would be reasonable to take account of quantum and statistical fluctuations.

While the time of the functional integration variables
in Eq.~\eqref{Eq:ReplicaAve} is (implicitly) assumed to be the same as that
for the observable, these times can be different.
Since the classical time evolution of canonical variables
is the canonical transformation
and the Hamiltonian $\Hml$ is a constant of motion on the classical path,
the integration measure is the same,
$\mathcal{D}\pi(t)\mathcal{D}\phi(t)=\mathcal{D}\pi_\mathrm{in}\mathcal{D}\phi_\mathrm{in}$ with $(\pi_\mathrm{in},\phi_\mathrm{in})$
being the initial field variables,
and the statistical weight is also the same,
$\exp(-\Hml(\phi(t),\pi(t))/\xi)=\exp(-\Hml(\phi_\mathrm{in},\pi_\mathrm{in})/\xi)$.
Thus the thermal average can be regarded
as the ``initial replica configuration average'',
provided that the initial replica ensemble is sampled
according to the statistical weight and the number of samples is large enough,
\begin{align}
\VEV{\mathcal O(\phi_\bx)}_t
=&
\frac{1}{\Zpf_R(\xi)}
\int \mathcal{D}\pi_\mathrm{in} \mathcal{D}\phi_\mathrm{in}\,
\widetilde{\mathcal{O}}(\phi_\bx(t,\pi_\mathrm{in},\phi_\mathrm{in}))
e^{-\Hml/\xi}
\nonumber\\
\simeq& \frac{1}{N_\mathrm{conf}}
\sum_{i=1}^{N_\mathrm{conf}}\widetilde{\mathcal{O}}
(\phi_\bx^{(i)}(t,\pi^{(i)}_\mathrm{in},\phi^{(i)}_\mathrm{in}))
\ .
\label{Eq:ReplicaAveMC}
\end{align}
We adopt this prescription in the later discussions.

\subsection{Partition function of Free Field}

Let us discuss 
the equilibrium property of replicas of the free field ($\lambda=0$),
where one can obtain the partition function analytically.
The Hamiltonian Eq.~\eqref{Eq:ReplicaH} is represented
by the Fourier components,
\begin{align}
\Hml^{(\lambda=0)}
=& \sum_{\bm{k},n}\frac12\left[\pi_{\nk}^2+\omega_{\nk}^2 \phi_{\nk}^2\right]
\ ,\label{Eq:FreeH}\\
\begin{pmatrix}
\phi_{\nk} \\
\pi_{\nk}
\end{pmatrix}
=&\frac{1}{\sqrt{NL^3}}\sum_{\bx,\tau}
\left[
e^{-i\bm{k}\cdot\bx+i\omega_n\tau}
\right]
\begin{pmatrix}
\phi_{\tx} \\
\pi_{\tx}
\end{pmatrix}
\ ,\\
\omega_{\bm{k}}^2
=&m^2+\bar{\bm{k}}^2
\ ,\quad
\bar{\bm{k}}^2\equiv 4\sum_{i=1}^D\sin^2(k_i/2)
\ ,\quad
\omega_{\nk}^2
=\omega_{\bm{k}}^2+4\xi^2\sin^2(\omega_n/2)
\ .\label{Eq:omegak}
\end{align}
Lattice momentum and the Matsubara frequency are defined as
$\bm{k}=(k_1,k_2,k_3)$
($k_i=2\pi m/L$, $m=0, 1, \ldots L-1$)
and 
$\omega_n=2\pi{n}/N$ ($n=0, 1, \dots N-1$), respectively.
Then the partition function is given by the Gaussian integral,
$\Zpf_{R}^{(\lambda=0)}=\prod_{\bm{k},n}\left(\xi/\omega_{\nk}\right)$,
where the integration measure is specified as
$\mathcal{D}\pi\mathcal{D}\phi
=\prod_{\bx,\tau}d\pi_{\tx}d\phi_{\tx}/(2\pi)
=\prod_{\bm{k},n}d\pi_{\nk}d\phi_{\nk}/(2\pi)$.
By using the Matsubara frequency summation formula
explained in Appendix~\ref{Sec:Matsubara},
the logarithm of the partition function for each lattice momentum $\bm{k}$
is found to be
\begin{align}
-\log\Zpf^{(\lambda=0)}_{\bm{k}}
=&\sum_n \log(\omega_{\nk}/\xi)
=\frac12\sum_n \log\left[
\left(\frac{\omega_{\bm{k}}}{2\xi}\right)^2+\sin^2(\omega_n/2)
\right]
+N \log 2
\nonumber\\
=&\log\left[2\sinh\left(\frac{\Omega_{\bm{k}}}{2T}\right)\right]
\ ,
\label{Eq:Zfree}
\end{align}
where $\Omega_{\bm{k}}$ is given as
\begin{align}
\Omega_{\bm{k}}
&=2\xi\,\mathrm{arcsinh}\,(\omega_{\bm{k}}/2\xi)
\ .
\end{align}
Now the partition function reads
\begin{align}
-\log\Zpf^{(\lambda=0)}
=&\sum_{\bm{k}}\log\left[
2\sinh\left(\frac{\Omega_{\bm{k}}}{2T}\right)
\right]
\ .
\end{align}
The energy expectation value for each momentum $\bm{k}$ is obtained as,
\begin{align}
\VEVev{E_{\bm{k}}^{(\lambda=0)}}
=&-\frac{\partial}{\partial\beta}\log\Zpf_{\bm{k}}^{(\lambda=0)}
=
\frac{1}{\sqrt{1+(\omega_{\bm{k}}/2\xi)^2}}
\left(
\frac{\omega_{\bm{k}}}{2}+\frac{\omega_{\bm{k}}}{e^{\Omega_{\bm{k}}/T}-1}
\right)
\ ,\label{Eq:BE}
\end{align}
where we have used the relation $\xi=NT$ and
$\VEVev{\cdots}$ denotes the thermal expectation value
at temperature $T$.
In the low frequency limit, $\omega_{\bm{k}}/T \ll 1$,
this energy converges to the classical value,
$\VEVev{E_{\bm{k}}^{(\lambda=0)}} \to T$.
The factor in front of the parentheses converges
to unity in the large $N$ limit,
$\xi=N T \to \infty$.
The first term in the parentheses is the zero point energy,
and should be subtracted in field theories.
The second term in the parentheses represents the thermal energy.
Compared with the classical field theory,
the Bose-Einstein distribution function appears
and the high-momentum components are exponentially suppressed
in the thermal part of energy.

\subsection{Time evolution of Free Field}
\label{Sec:Free}

Time evolution of phase space variables of the free field
in the momentum representation are obtained as
\begin{align}
\phi_\nk(t)=&\phi_\nk(0)\cos\omega_\nk{t}
+\frac{\pi_\nk(0)\sin\omega_\nk t}{\omega_\nk}
\ ,\label{Eq:phinkt}\\
\pi_\nk(t)=&-\omega_\nk\phi_\nk(0)\sin\omega_\nk{t}
+\pi_\nk(0)\cos\omega_\nk t
\ .\label{Eq:pinkt}
\end{align}
By using Eqs.~\eqref{Eq:phinkt} and \eqref{Eq:pinkt},
we can evaluate spacetime dependence of
the field variables and the two point functions,
\begin{align}
\phi_\tx(t)=&
\frac{1}{\sqrt{NL^3}}
\sum_\nk e^{i\bm{k}\cdot\bm{x}-i\omega_n\tau}\phi_\nk(t)\ ,\\
\VEVev{\phi_{\bm{x}}(t)\phi_{\bm{y}}(t')}
\equiv&
\frac{1}{N}\sum_\tau \VEVev{\phi_{\tx}(t)\phi_{\tau\bm{y}}(t')}
\nonumber\\
=&\frac{1}{N^2L^3}\sum_{\tau,n,\bm{k},n',\bm{k}'}
e^{
i\bm{k}\cdot\bm{x}-i\omega_n\tau
-i\bm{k}'\cdot\bm{y}+i\omega_{n'}\tau}
\VEVev{\phi_{\nk}(t)\phi^*_{n'\bm{k}'}(t')}
\ .
\end{align}
The thermal ensemble average for $\phi_\nk(0)$ and $\pi_\nk(0)$,
whose distributions are Gaussians, is taken as
\begin{align}
\VEVev{\phi_{\nk}(0)\phi^*_{n'\bm{k}'}(0)}=\frac{\xi}{\omega_\nk^2}
\,\delta_{n,n'}\,\delta_{\bm{k},\bm{k}'}
\ ,\quad
\VEVev{\pi_{\nk}(0)\pi^*_{n'\bm{k}'}(0)}=\xi
\,\delta_{n,n'}\,\delta_{\bm{k},\bm{k}'}
\ .
\end{align}
Then we find that the two point functions are given as
\begin{align}
\VEVev{\phi_{\nk}(t)\phi^*_{n'\bm{k}'}(t')}=&\frac{\xi}{\omega_\nk^2}
\,\delta_{n,n'}\,\delta_{\bm{k},\bm{k}'}\,\cos\left\{\omega_\nk(t-t')\right\}
\ ,\\
\VEVev{\phi_{\bm{x}}(t)\phi_{\bm{y}}(t')}
=&\frac{1}{L^3}\sum_{n,\bm{k}}
\frac{T}{\omega_\nk^2}
e^{i\bm{k}\cdot(\bm{x}-\bm{y})}\cos\left\{\omega_\nk(t-t')\right\}
\ .
\end{align}
In the later discussions, the following two point functions
will be used and discussed,
\begin{align}
\Delta 
=& \VEVev{\phi^2}
=\frac{1}{NL^3}\sum_{\tau,\bx}\VEVev{\phi_\tx(t)\phi_\tx(t)}
=\frac{1}{NL^3}\sum_{n,\bk}\VEVev{\phi_\nk(t)\phi^*_\nk(t)} 
\nonumber\\
=&\frac{1}{L^3}\sum_{n,\bm{k}}
\frac{T}{\omega_\nk^2}
=\frac{1}{L^3}
\sum_{\bm{k}}
\frac{1}{\omega_{\bm{k}}\sqrt{1+(\omega_{\bm{k}}/2\xi)^2}}
\left[
\frac12+\frac{1}{e^{\Omega_{\bm{k}}/T}-1}
\right]
\ ,\label{Eq:Delta}\\
C(t)
=& \frac{1}{L^3} \sum_{\bm{x},\bm{y}}
\VEVev{\phi_{\bm{x}}(t_0+t)\phi_{\bm{y}}(t_0)}
= \frac{1}{NL^3} \sum_{\tau,\bm{x},\bm{y}}
\VEVev{\phi_{\tau\bm{x}}(t_0+t)\phi_{\tau\bm{y}}(t_0)}
\nonumber\\
=& \frac{1}{NL^3} \sum_{\tau,\bm{x},\bm{y}}
\VEVev{\phi_{\tau\bm{x}}(t_0+t)\phi_{\tau\bm{y}}(t_0)}
=\sum_{n}
\frac{T}{\omega^2_{n\bm{0}}}
\cos\omega_{n\bm{0}}t
\ .
\label{Eq:GF}
\end{align}
The first one ($\Delta=\VEVev{\phi^2}$) appears in the one-loop diagram
and diverges in the continuum limit.
The second one ($C(t)$) is the unequal-time two-point function at zero momentum,
referred to as the the time-correlation function in the later discussions,
and is expected to show oscillatory behavior with frequency of the thermal mass.
In equilibrium, the ``trigger'' time of the measurement, $t_0$,
can be taken arbitrary.

\subsection{Mass renormalization}

\begin{figure}[htbp]
\begin{center}
\includegraphics[width=5cm,bb=40 320 460 550,clip]{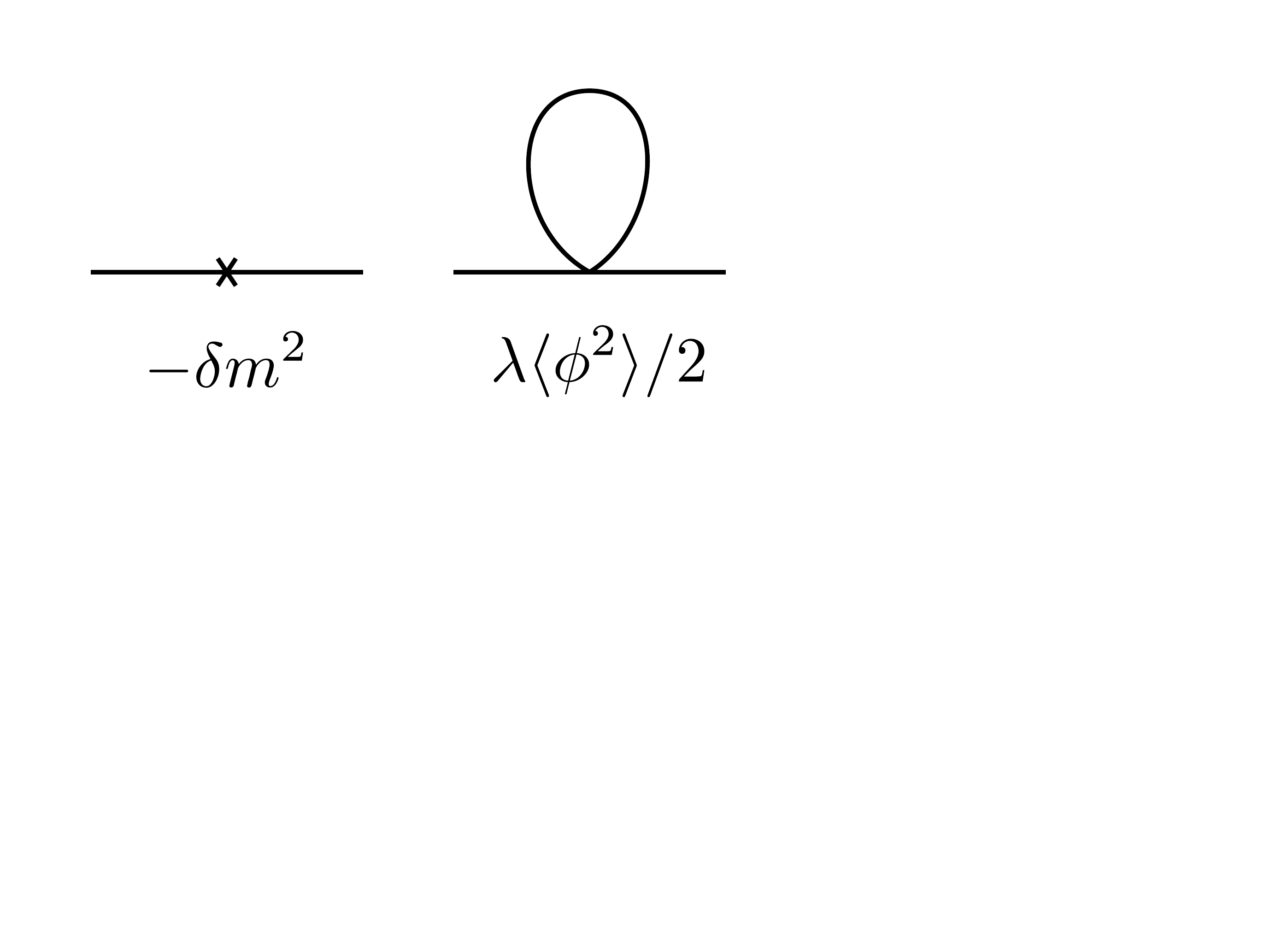}
\end{center}
\caption{Mass counterterm and the one-loop diagram
contributing to the thermal mass.}
\label{Fig:One_loop}
\end{figure}

In the replica evolution with finite coupling,
we need to take care of the mass renormalization
as in the standard treatment of quantum field theory.
We consider the contribution of the one-loop diagram and the counterterm
shown in Fig.~\ref{Fig:One_loop},
then the thermal mass including the contribution from the interaction
is found to be
\begin{align}
M^2
=&m^2-\delta{m}^2+\frac{\lambda}{2}\VEVev{\phi^2}
=m^2-\delta{m}^2+\frac{\lambda\Delta}{2}
=m^2 + \frac{\lambda}{2}\VEV{\phi^2}_\mathrm{ren}
\ ,\label{Eq:mass}\\
\delta{m}^2
=&\frac{\lambda}{2}\VEV{\phi^2}_\mathrm{div}
=\frac{\lambda}{2}\frac{1}{L^3}\sum_{\bm{k}}
\frac{1}{2\omega_{\bm{k}}\sqrt{1+(\omega_{\bm{k}}/2\xi)^2}}
\ ,\label{Eq:dmsq}\\
\VEV{\phi^2}_\mathrm{ren}
=&\VEVev{\phi^2}-\VEV{\phi^2}_\mathrm{div}
=\frac{1}{L^3}
\sum_{\bm{k}}
\frac{1}{\omega_{\bm{k}}\sqrt{1+(\omega_{\bm{k}}/2\xi)^2}}
\frac{1}{e^{\Omega_{\bm{k}}/T}-1}
\ .\label{Eq:phisq_ren}
\end{align}
We choose the counterterm $\delta m^2$ 
so that it cancels the divergent contribution to the mass,
$\lambda\VEV{\phi^2}_\mathrm{div}/2$,
where $\VEV{\phi^2}_\mathrm{div}$
is the divergent part of $\Delta=\VEVev{\phi^2}$
given in the first term in the bracket in Eq.~\eqref{Eq:Delta}.
The mass term induced by the interaction, $\lambda\VEVev{\phi^2}/2$, 
coincides with the factorization (Wick contraction) of the interaction term
appearing in the equation of motion,
$\lambda\phi^3/6 \simeq \lambda\VEVev{\phi^2}\phi/2$.

\begin{figure}[htbp]
\begin{center}
\includegraphics[width=10cm,bb=0 0 360 250]{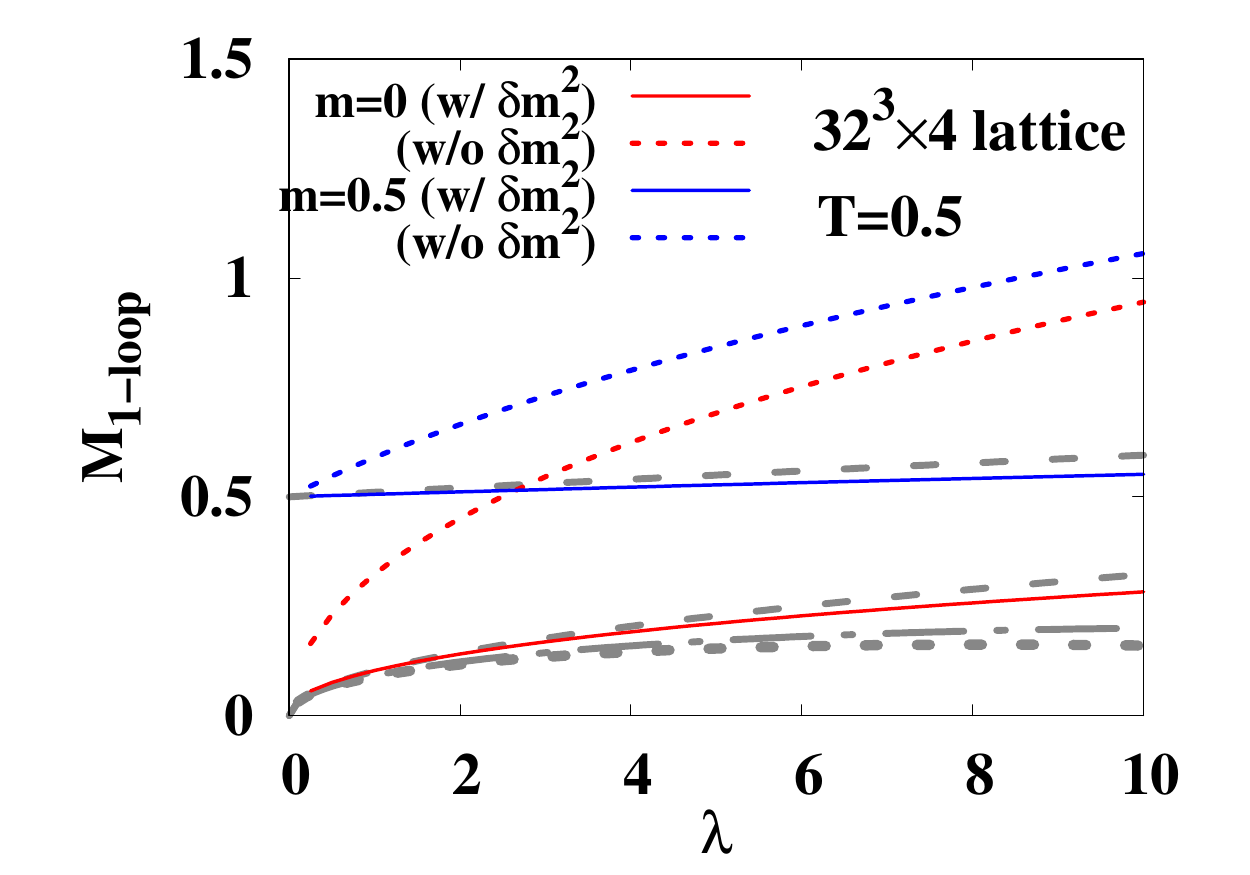}
\end{center}
\caption{Thermal mass in $\phi^4$ theory on the lattice. 
Red and blue curves show the results of the thermal mass,
$M=\omega(\bm{k}=\bm{0})$, with $m=0$ and $m=0.5$, respectively.
Dotted and solid curves show the results without and with
the counterterm, $\delta m^2$, respectively.
Long-dashed, dot-dashed and short-dashed lines show
the perturbative calculation results of the thermal mass 
in the continuum limit 
with the leading order ($M_\mathrm{LO}$),
the resummed one loop ($M_\mathrm{resum}$)
and the two loop ($M_\mathrm{2{\operatorname{-}}loop}$) effects, respectively.
}\label{Fig:dxsq}
\end{figure}

In Fig.~\ref{Fig:dxsq}, we show the thermal mass
calculated on a $32^3\times 4$ lattice at $T=0.5$.
Thermal mass $M$ is obtained by solving Eqs.~\eqref{Eq:mass}, \eqref{Eq:dmsq}
and \eqref{Eq:phisq_ren} self-consistently
by taking account of $M$ dependence of $\omega_{\bm{k}}$ and $\Omega_{\bm{k}}$.
For comparison, we also show the results of the leading order estimate
in the continuum and large $N$ limit,
$1/a \to \infty$, $L\to\infty$ and $N\to\infty$~\cite{Kapusta},
\begin{align}
M_\mathrm{LO}^2=m^2+\lambda T^2/24 .
\end{align}
For $m=0$, we also show the results with resummed one-loop contribution
from the self-consistent treatment~\cite{Kapusta,TwoLoop},
\begin{align}
M^2_\mathrm{resum}
=\frac{\lambda T^2}{24} \left[1-\frac{3}{\pi}\sqrt{\frac{\lambda}{24}}\right]
\ ,
\end{align}
and the two-loop calculation result~\cite{TwoLoop},
\begin{align}
M^2_\mathrm{\TL}
=&\frac{\lambda T^2}{24}\left\{
 1-\frac{3}{\pi}\sqrt{\frac{\lambda}{24}}
 +\frac{\lambda}{(4\pi)^2}\left[
    \frac{3}{2}\log\left(\frac{T^2}{4\pi\mu^2}\right)
   +2\log\left(\frac{\lambda}{24}\right)+\alpha
  \right]
 \right\}
\ ,
\end{align}
where $\alpha=8.8865\ldots$, calculations is carried out at negligible $m/T$,
and we take the renormalization scale as $\mu=2\pi T$.
The obtained thermal mass on the lattice at $m=0$ is
between $M_\mathrm{LO}$ and $M_\mathrm{resum}$.
The deviation from $M_\mathrm{resum}$ may be due to the limited momentum range
and the additional factor of $1/\sqrt{1+(\omega_{\bm{k}}/2\xi)^2}$
in $\VEV{\phi^2}_\mathrm{ren}$ shown in Eq. \eqref{Eq:phisq_ren}.

\section{Numerical results of replica evolution in scalar field theory}
\label{Sec:Results}

We shall now numerically evaluate the time evolution of replicas
of classical field by using the replica Hamiltonian Eq.~\eqref{Eq:ReplicaH}
defined in the 4D spacetime including the imaginary time.
In order to examine the validity of replica evolution,
we discuss the time-correlation function $C(t)$,
which is the unequal-time two-point function at zero momentum.
From the time correlation $C(t)$ obtained by the replica evolution, we extract the thermal mass $M$ and the damping rate $\gamma$, and compare them with the perturbative estimates.

\subsection{Setup}

We show the numerical results of time evolution of replicas
on a $32^3\times 4$ lattice ($L=32, N=4$)
at $T=0.5$ in the coupling range of $0.5 \leq \lambda \leq 10$
with $m=0$ and $m=0.5$, as an example.
The equation of motion is solved in the leap-frog method
until $t=500$ with the time step of $\Delta t=0.025$
after the equilibration described below.
When we take account of the mass renormalization,
we subtract the divergent part of the induced mass by using
the one-loop calculation results given in Eq.~\eqref{Eq:dmsq}.

We prepare the initial condition by using the Langevin equation
at the replica temperature of $\xi=NT=2$,
\begin{align}
\frac{d\pi_\tx}{dt}=\frac{\partial \Hml}{\partial\phi_\tx}
-\Gamma \pi_\tx + \sqrt{2\Gamma\xi}\,\zeta_\tx(t)
\ ,
\end{align}
where the drift constant is taken to be $\Gamma=0.5$
and $\zeta_\tx(t)$ is the white noise, 
$\VEV{\zeta_\tx(t)\,\zeta_{\tau'\bm{x}}(t')}=
\delta_{\tau,\tau'}\,\delta_{\bm{x},\bm{x}'}\,\delta(t-t')$.
Since the relation of $\VEVev{\pi_\tx^2}=\xi$ should be satisfied in equilibrium,
we rescale $\pi$ field at each step of the Langevin evolution.
The equilibration time to prepare the initial condition 
is set to be $t_\mathrm{eq}=20$ in the lattice unit,
which is found to be reasonably long in the coupling range
for $\lambda \geq 4$.
At smaller coupling, we take $\tau_\mathrm{eq}=100, 60$ and $40$
for $\lambda=0.5, 1$ and $2$, respectively.

It should be noted that any initial condition can be used, in principle,
as long as the $\Hml$ distribution in the ensemble is consistent
with that in equilibrium.
Provided that the system is chaotic, 
all the phase-space points having the given $\Hml$ value
are sampled in a long-time evolution~\cite{Poincare}.
However, it generally takes more time for a Hamiltonian system
to reach the equilibrium than in the Langevin equation.
For example,
the time needed to achieve equilibrium using the Langevin equation
is found to be much shorter than the intrinsic relaxation time
of the classical scalar field ($N=1$) in Ref.~\cite{Matsuda}.
Thus it is expected that equilibrium configurations are efficiently
obtained by using the Langevin equation.

We evaluate the time-correlation function $C(t)$
of the zero momentum component of the field variable,
$\phi^{(i)}_{\tau,\bm{k}=\bm{0}}(t)=\sum_{\bm{x}}\phi^{(i)}_\tx(t)/\sqrt{L^3}$,
with $i$ being the configuration index.
The equilibrium average of the correlation function is obtained
as the average over the replica ensemble,
as shown in Eq.~\eqref{Eq:GF},
where the replica ensemble is prepared by the Langevin equation discussed above.
The number of replica configurations
is taken to be $N_\mathrm{conf}=1000$.
In order to reduce the statistical error, we also take average
over the trigger time $t_0$.

With this setup, we have solved the time evolution of replica
configurations, where the ensemble average should be consistent
with the equilibrium expectation value.
The ensemble average of $\pi_\tx^2$ is found to be $(0.1-0.2)\,\%$
larger than $\xi$ in the present setup.
The overestimate can be suppressed
with smaller drift coefficient and longer equilibration time,
but it takes more time for the calculation
and the above deviation would be small enough.

\subsection{Momentum distribution and Rayleigh-Jeans divergence}

\begin{figure}[htbp]
\begin{center}
\includegraphics[width=7.5cm,bb=0 0 360 250]{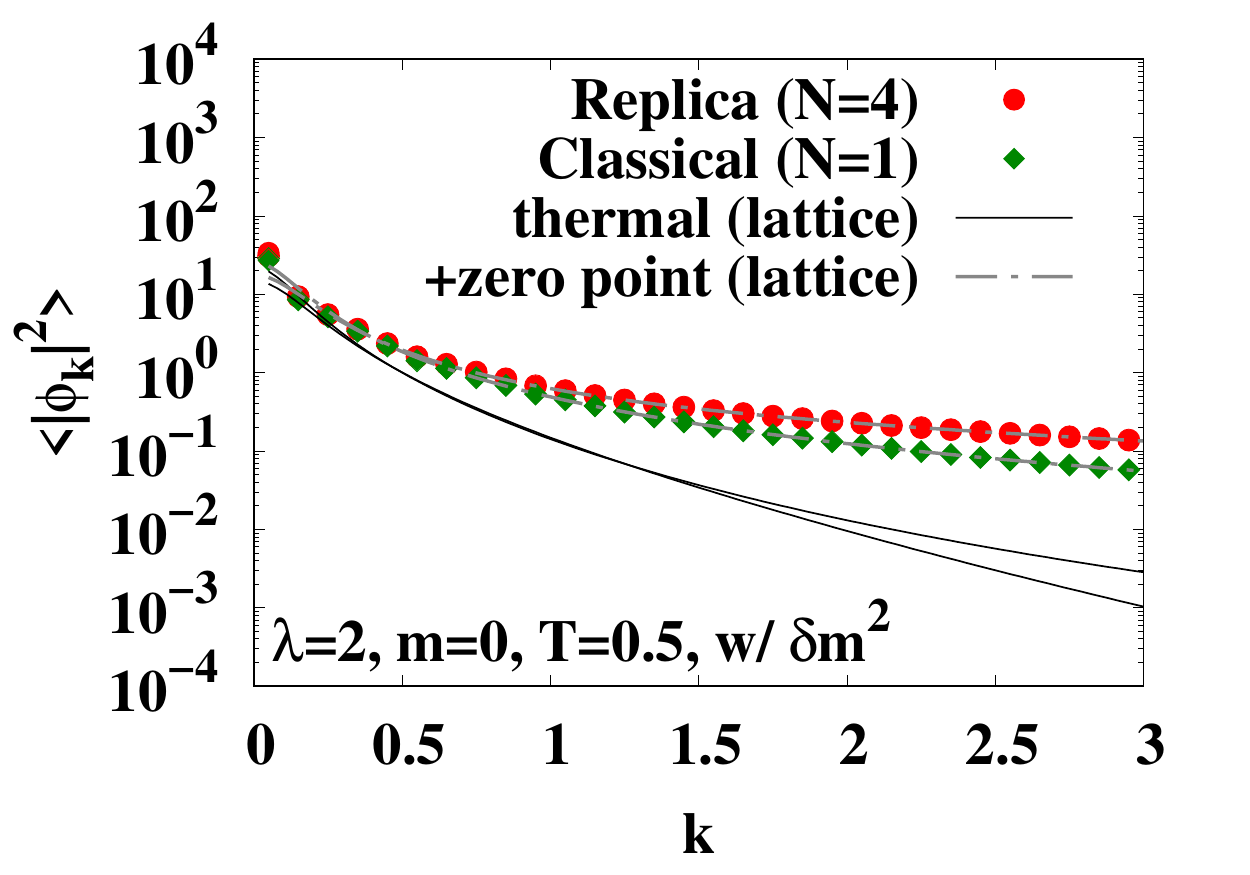}%
\includegraphics[width=7.5cm,bb=0 0 360 250]{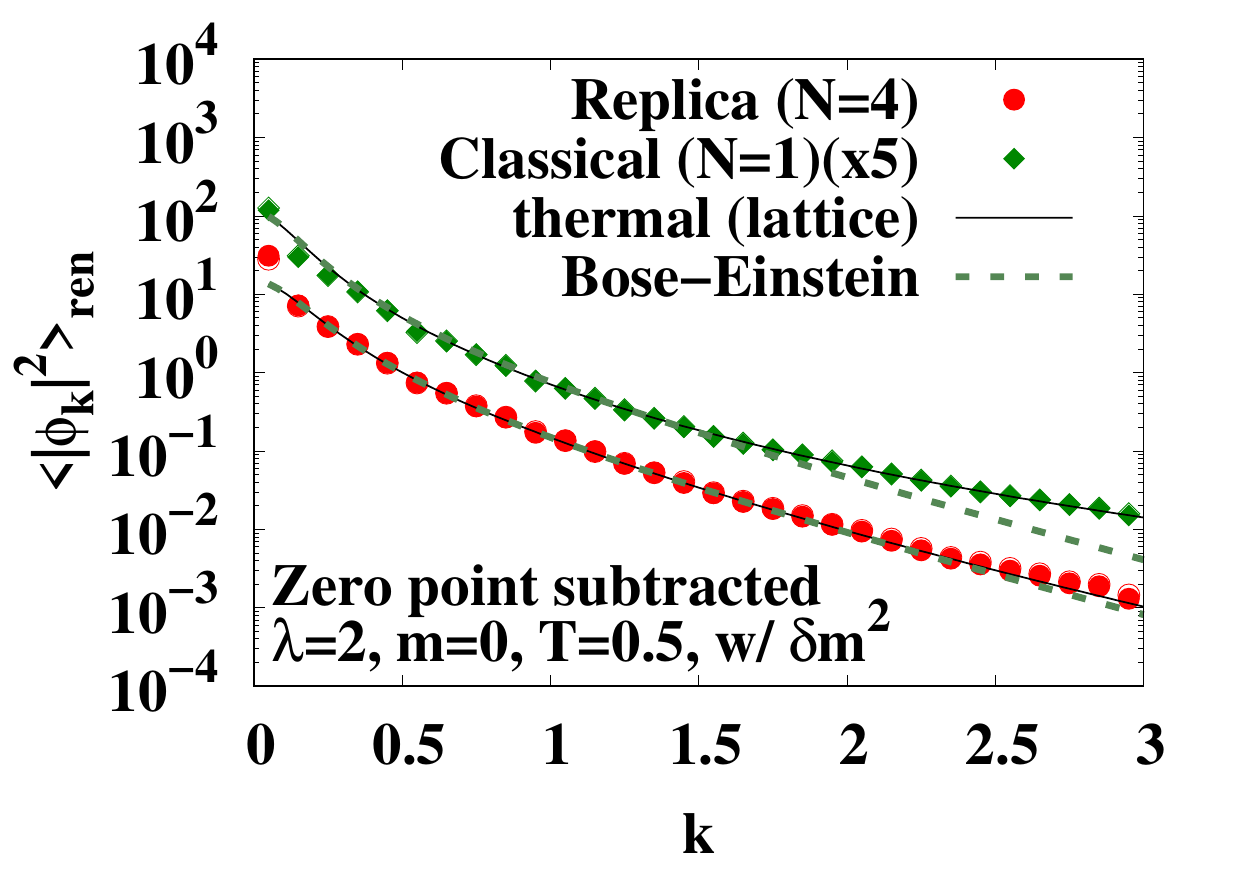}%
\end{center}
\caption{Momentum distribution $\VEVev{|\phi_{\bm{k}}|^2}$
obtained in replica ($N=4$, circles)
and classical field ($N=1$, diamonds) ensembles at $m=0$ and
$\lambda=2$ on the $32^3$ lattice
in comparison with the thermal part of the distribution
in the free field (solid curves).
Filled and open symbols show the results
at $t=t_\mathrm{eq}$ and $t=t_\mathrm{eq}+500$, respectively.
Left panel shows the results including the zero point part,
whose sum over the momenta diverges.
Solid curves show the thermal part, 
and dash-dotted curves show the results including the zero point part.
Right panel shows the results of the finite (renormalized) part of the
momentum distribution, where the zero point part is subtracted.
Solid curves show the thermal contribution on the lattice,
and dashed curves show their large $N$ limit, $N\to\infty$, which is 
equivalent to the Bose-Einstein distribution multiplied by $1/\omega_{\bm{k}}$.
}\label{Fig:phiksq}
\end{figure}

Before discussing real time evolution,
let us take a look at the thermal expectation value
of the momentum distribution,
\begin{align}
\VEVev{|\phi_\bk|^2}
=&\frac{1}{N}\sum_{\tau}\VEV{\phi_{\tau\bk}\phi^*_{\tau\bk}}
=\frac{1}{N}\sum_{n}\VEV{\phi_{\nk}\phi^*_{\nk}}
\ ,\label{Eq:pdist}
\end{align}
as a function of momentum $k=(\bar{\bm{k}}^2)^{1/2}$.
This appears in $\Delta=\VEV{\phi^2}$ (Eq.~\eqref{Eq:Delta})
and also in the energy in the form of 
$\omega_\bk^2\VEVev{|\phi_\bk|^2}=(m^2+k^2)\VEVev{|\phi_\bk|^2}$
in Eq.~\eqref{Eq:FreeH}.
In the free field case, the momentum distribution is given as
\begin{align}
\VEVev{|\phi_\bk|^2}
=&\frac{1}{\omega_{\bm{k}}\sqrt{1+(\omega_{\bm{k}}/2\xi)^2}}
\left[
\frac12+\frac{1}{e^{\Omega_{\bm{k}}/T}-1}
\right]
\ ,\label{Eq:pdistfree}
\end{align}
as discussed in Sec.~\ref{Sec:Free}.
The first term in the bracket in Eq.~\eqref{Eq:pdistfree}
shows the zero point energy contribution which should be subtracted,
and the second term shows the thermal part of the momentum distribution
on the lattice which converges to the Bose-Einstein distribution function
$1/[\exp(\omega_{\bm{k}}/T)-1]$ in the large $N$ limit.
In Fig.~\ref{Fig:phiksq}, we show
the replica ($N=4$) and classical field ($N=1$) ensemble results at $m=0$ and $\lambda=8$.
We here adopt the thermal mass evaluated from the time-correlation function
discussed later.
These results agree with the free field results on the lattice.
When the divergent part is subtracted (right panel),
the momentum distributions approximately show exponential decay,
as the Bose-Einstein distribution does.
It may be interesting to find that even in the case of classical field,
an approximate exponential decay is found,
while the thermal part in the free field 
on the lattice deviate those in the large $N$ limit at high momenta, $k>1.5$.
This approximate exponential decay comes from the decomposition
of the divergent and finite parts as shown in Eq.~\eqref{Eq:pdistfree}.
But this is not the end of the story.

\begin{figure}[htbp]
\begin{center}
\includegraphics[width=10cm,bb=0 0 360 250]{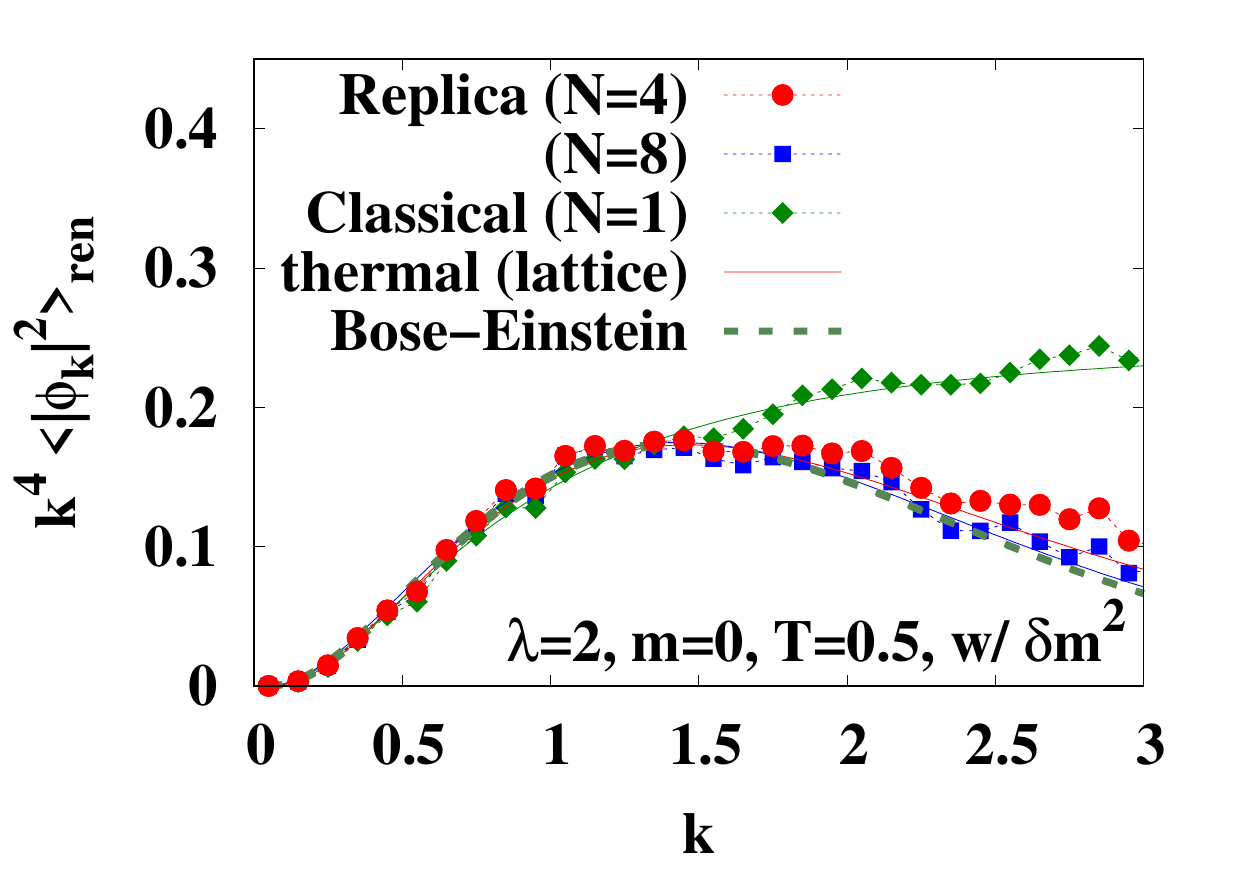}%
\end{center}
\caption{Renormalized momentum distribution multiplied by $k^4$,
$k^4\VEVev{|\phi_{\bm{k}}|^2}_\mathrm{ren}$,
obtained in replica ($N=4$, circles and $N=8$, squares)
and classical field ($N=1$, diamonds) ensembles on the $32^3$ lattice
at $m=0$, $\lambda=2$, and $T=0.5$.
Solid curves show the thermal part of the distribution in the free field
on the lattice, and the dashed curve shows the large $N$ limit,
corresponding to the Bose-Einstein distribution
multiplied by $k^4/\omega_{\bm{k}}$.
}\label{Fig:RJ}
\end{figure}

Next, let us discuss the Rayleigh-Jeans divergence.
The momentum distribution appears in energy in the form of
$k^2\VEVev{|\phi_\bk|^2}$ and the number of momentum modes
increase with the momentum as $k^2$.
Thus the energy contains the kinetic energy part of
$\int dk k^4 \VEVev{|\phi_\bk|^2} / 2\pi^2$ in the continuum limit.
In Fig.~\ref{Fig:RJ}, we show the momentum distribution multiplied by $k^4$.
We note that the classical field results ($N=1$)
seem to saturate to a constant value.
This behavior can be understood from the decomposition in Eq.~\eqref{Eq:pdistfree}.
The thermal part in the decomposition in Eq.~\eqref{Eq:pdistfree}
is not necessarily exponentially suppressed 
but rationally suppressed at large $k$.
Because of the functional form,
$\mathrm{arcsinh}\,x=\log(\sqrt{1+x^2}+x)\simeq \log(2x)$ at large $x$,
the ``exponential'' reads
$\exp(-\Omega_\bk/T)\simeq\exp(-2N\log(\omega_\bk/NT))=(\omega_\bk/NT)^{-2N}$
for large $\omega_\bk$, $\omega_\bk \gg \xi$.
Then for high momentum, $k \gg m$ and $k \gg NT$, we find
\begin{align}
\VEV{|\phi_\bk|^2}
\simeq&
\frac{2NT}{k^2}\exp(-\Omega_\bk/T)
\to
2(NT)^{2N+1}
k^{-2(N+1)}
\ ,\\
k^4\,\VEV{|\phi_\bk|^2}\to&
\,
2(NT)^{2N+1}k^{-2(N-1)}
\ .
\end{align}
Then $k^4\VEV{|\phi_\bk|^2}$ converges to a constant with $N=1$.
By substituting $T=0.5$ and $N=1$, the asymptotic value is found to be
$2(NT)^{2N+1}=0.25$,
which is close to the classical field results at large $k$.
Thus we cannot fully remove the Rayleigh-Jeans divergence in the classical field
in the present subtraction scheme without additional matching procedure.

In contrast,
the replica results ($N=4$ and $N=8$) of the momentum distribution
show suppressed behavior at high momentum
as the Bose-Einstein distribution does,
and $k^4\VEV{|\phi_\bk|^2}$ also decreases at large $k$.
For the convergence of energy, $k^4\VEV{|\phi_\bk|^2}$ needs to converge to zero
faster than $1/k$, then we find the constraint on $N$ as
$2(N-1) > 1$, or $N > 3/2$.
Thus we can expect that the integral would converge to a finite value
also in the continuum limit,
and the Rayleigh-Jeans divergence in energy can be fully removed
in the replica evolution even with $N=4$.
With $N=8$, the momentum distribution is closer to the Bose-Einstein
distribution in the large $N$ limit.

\subsection{Time-correlation function, thermal mass and damping rate}

\begin{figure}[htbp]
\begin{center}
\includegraphics[width=15cm,bb=0 0 237 165]{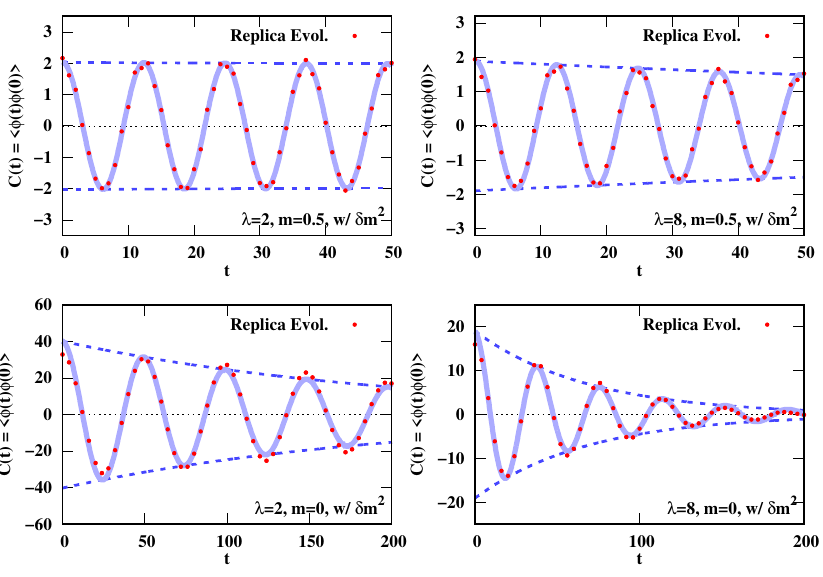}%
\end{center}
\caption{Time-correlation function $C(t)$ in replica evolution
on the $32^3\times4$ lattice with $\lambda=2$ (left) and $8$ (right)
with mass renormalization.
Upper and lower panels show the results at $m=0.5$ and $m=0$, respectively.
Replica evolution results (circles)
are compared with the fitting results with $f_{1\omega}$
(thick blue curves).
}\label{Fig:GF}
\end{figure}

We now proceed to discuss the time-correlation function $C(t)$
obtained from the time evolution of replica ensemble.
In Fig.~\ref{Fig:GF}, we show the time-correlation function $C(t)$
from replica evolution at $\lambda=2$ and $8$ with mass renormalization.
We find that the time-correlation function is well described
by the single damped oscillator $f_{1\omega}(t)$,
\begin{align}
f_{1\omega}(t)=A \exp(-\gamma t)\,\cos(Mt+\delta)\ .
\label{Eq:1w}
\end{align}
as shown by the thick blue curves.

\begin{figure}[htbp]
\begin{center}
\includegraphics[width=15cm,bb=0 0 360 126]{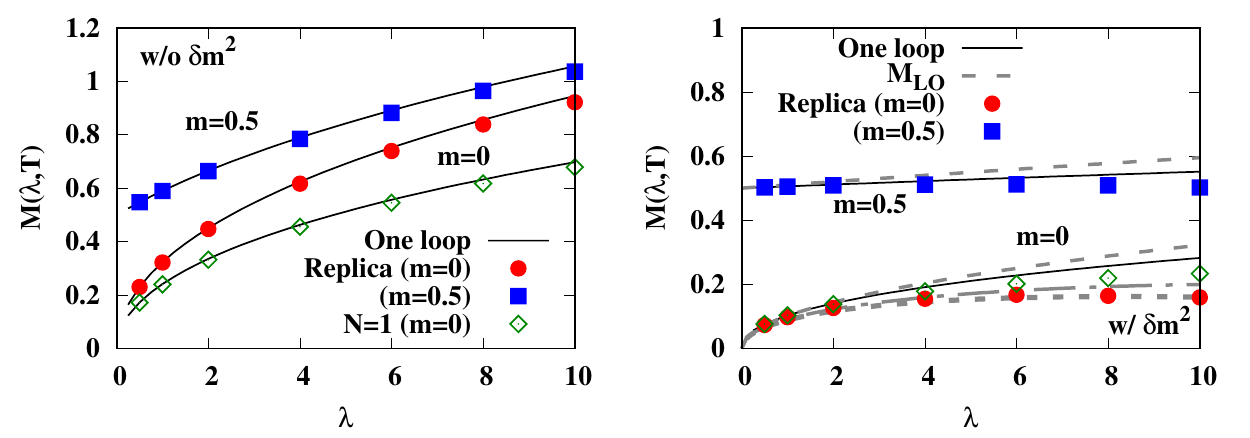}%
\end{center}
\caption{Thermal mass $M$ obtained as the frequency
of the time-correlation function $C(t)$ in replica evolution
without (left) and with (right) mass counterterm at $m=0$ and $m=0.5$
(filled circles and squares).
Diamonds show the results of classical field evolution ($N=1$) at $m=0$.
Long-dashed, dot-dashed and short-dashed lines in the right panel show
the perturbative calculation results
with the leading order ($M_\mathrm{LO}$),
the resummed one loop ($M_\mathrm{resum}$)
and the two loop ($M_\mathrm{2{\operatorname{-}}loop}$) effects, respectively.
}\label{Fig:omg}
\end{figure}

We have obtained the thermal mass $M$ and the damping rate $\gamma$
by fitting the parameters $(M,\gamma,A,\delta)$ in $f_{1\omega}$
to $C(t)$ obtained from the replica evolution.
In Fig.~\ref{Fig:omg},
we show the coupling dependence of the thermal mass $M$
obtained from the time-correlation function $C(t)$ in replica evolution
without (left) and with (right) mass renormalization.
We show the fitting results using the single oscillator function $f_{1\omega}$.
The fitting results 
are close to the one loop calculation results without mass renormalization.
With mass renormalization,
the thermal mass is consistent with the one loop results at small coupling
but considerably smaller than the one loop results
in the strong coupling region.
Replica evolution results at $m=0$ agree with the two loop results,
while the classical field results ($N=1$) at $m=0$ show weaker reduction
from the leading order results.
Thus the replica evolution is expected to give higher order interaction effects
over the one loop.
For more serious comparison,
we need to take account of two-loop counterterms of mass and coupling,
and to choose the renormalization scale $\mu$ consistent with the present
lattice calculation, but these are beyond the scope of this work.

\begin{figure}[htbp]
\begin{center}
\includegraphics[width=15cm,bb=0 0 360 115]{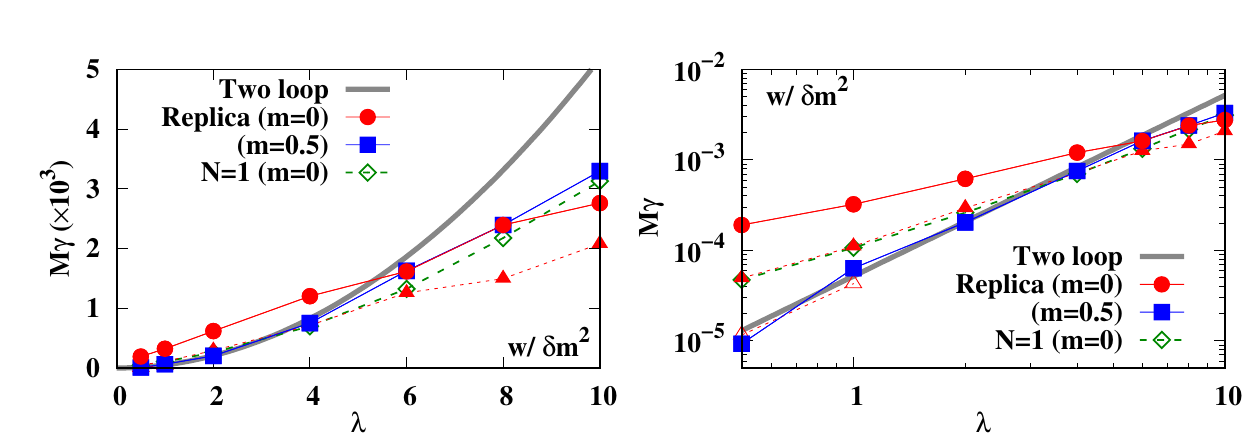}%
\end{center}
\caption{Damping rate $\gamma$ multiplied by the mass $M$
obtained from the time-correlation function
$C(t)$ in replica evolution
with mass counterterm at $m=0$ and $m=0.5$ (circles and squares)
in comparison with the two-loop calculation results (solid curves).
Open and filled triangles show the results at $m=0$
obtained from the analyses 
of the Fourier transform of the time-correlation function
using $\rho_\Delta$ and $f_{2\omega}$,
respectively, as discussed in Appendix \ref{Sec:spec}.
Open diamonds show the classical field results ($N=1$) at $m=0$.
}\label{Fig:gam}
\end{figure}

In Fig.~\ref{Fig:gam},
we show the damping rate $\gamma$ as a function of the coupling.
We compare the replica evolution results with mass counterterm in comparison
with the two loop calculation results \cite{TwoLoop},
\begin{align}
\gamma = \frac{\lambda^2T^2}{1536\pi M}
\ ,\label{Eq:gamma_2-loop}
\end{align}
which is known to agree with the plasmon damping rate
after the matching to quantum theory
by substituting the leading order thermal mass estimate,
$M_\mathrm{LO}=\sqrt{\lambda T^2/24}$.
In this expression,
the product $M\gamma$ is found to be independent of the mass.

The replica evolution results with $m=0.5$
seem to roughly agree with the perturbation calculation results
in the small coupling region,
and start to deviate from the perturbative estimate at larger coupling,
$\lambda \geq 4$.
Deviations at $\lambda \geq 4$ would be due to
higher-order effects of the coupling, fluctuations,
or the higher-momentum components ignored in the lattice discretization.
With $m=0$, the damping rates are smaller than the perturbative estimate
in the strong coupling region, as in the $m=0.5$ cases.
At $\lambda \leq 4$, by comparison,
the damping rate tends to be larger than the perturbative estimates.
Since the thermal mass is small in this region of coupling,
$M=0.085$ and $0.12$ at $\lambda=0.5$ and $1$, respectively,
we may need larger size lattice.
The apparent larger damping rate at small thermal mass may be related with 
the fragmentation of the single particle mode into several modes.
The Fourier transform at small coupling and $m=0$
is found to be represented better
by the superposition of several damped oscillators,
each of which has a small width.
If we adopt the width for each of the modes as the damping rate,
the results at $\lambda=0.5$ and $1$ agree with the perturbative estimates
as shown by open triangles in Fig.~\ref{Fig:gam}.
The analysis of the Fourier transform is given in Appendix~\ref{Sec:spec}.
It should be also noted that the classical field results roughly agree
with the perturbative calculation,
as already noted in previous works~\cite{Aarts,Aarts:2001yn,Aarts-spec}.

Before closing this section, we would like to mention that
$C(t)$ is the time-symmetric part of the two-point function, $C(-t)=C(t)$,
whose transformation is referred to as the {\em statistical} function.
In quantum field theory, the spectral function is more important,
but we need to evaluate the commutator of unequal-time field variables
by using, for example, the Poisson bracket~\cite{Aarts-spec}.
We show the time-odd part of the time-correlation function 
in a harmonic oscillator in Appendix~\ref{Sec:PB},
but we leave evaluating the spectral function in field theories
in the future work.

\section{Summary and perspectives}
\label{Sec:Summary}

We have investigated the simultaneous real-time evolution
of several classical field configurations, referred to as replicas,
which interact with the nearest neighbor replicas
with a specific form of interaction, the $\tau$-derivative term.
Classical evolution of replica ensemble is found to provide 
the correct quantum field partition function in equilibrium.
The average of field variables over replica indices approximately obeys
the classical field equation of motion
when the fluctuations among the replicas are small.
The exponential suppression factor of high momentum modes
appears from the sum over the replica index,
which can be regarded as the imaginary time,
provided that the zero point energy contribution is subtracted.
We have examined the behavior of the time-correlation function
(the unequal-time two-point function)
at zero momentum without and with the mass counterterm.
The time-correlation function in the replica evolution
is expected to show oscillatory behavior with the frequency
of the thermal mass $M$
at temperatures $T/M \gtrsim 0.5$, as demonstrated in the quantum mechanics
of the harmonic oscillator.
The time-correlation function in the $\phi^4$ theory
seems to show reasonable behavior:
The thermal mass of the zero momentum mode roughly agrees
with the perturbative calculation results at small coupling.
Thus the replica evolution should be useful to describe
real-time behavior in equilibrium.

It would be desired to further examine the replica evolution
as a candidate of the frameworks to describe non-equilibrium real-time quantum-field evolution.
As long as the distribution of initial replica Hamiltonian values
is properly given, distribution of field variables in replica ensemble
should finally relax to the correct quantum statistical distribution,
even if one starts from far-from-equilibrium configurations.
In discussing non-equilibrium real-time evolution,
however, it would be necessary to introduce additional time scale.
It should be noted that 
replica evolution is conjectured to be useful in a heuristic context,
but it is not derived based on some principle.
Thus formal {\em derivation} or {\em justification} is desired.
For example, the equivalence between the classical field theory
and the Boltzmann equation~\cite{Mueller:2002gd}
would be a good guide for the formal discussions.
It is also interesting to discuss, for example,
the $O(N)$ model, where there exist results
of dynamical calculations using the two particle irreducible (2PI) 
effective action~\cite{Berges:2004yj,Aarts:2001yn}.

Once the present framework is proven to be useful
in describing quantum field evolution toward equilibrium,
application to the Yang-Mills field is another important subject to study.
The temporal component of the vector field is usually
Wick rotated in the imaginary time formalism
and we cannot apply the replica evolution as it is.
However, spatial components are the same in the imaginary
and real time formalism and we can apply the replica evolution.
Thus it is possible to examine the replica evolution of classical 
Yang-Mills field in the temporal gauge
where the temporal components of the vector field are set to be zero.
Then it is interesting to examine whether or not
the quantum statistical features affect the dynamical evolution
in the initial stage of high-energy heavy-ion collisions.
In order to describe inhomogeneous and nonequilibrium evolution,
it is necessary to take account of spacetime dependence of temperature,
which may need to invoke
the nonequilibrium statistical operator (NSO) method~\cite{NSO}.
In the NSO method, the inverse temperature and several other variables are
introduced as the local conjugate fields of the corresponding density fields,
and are determined by the time evolution and the initial condition
under the assumption that the local Gibbs distribution in the initial state.
Combining the NSO method and the replica evolution may be a challenging
but valuable subject to study.

\section*{Acknowledgments}
The authors would like to thank
J{\o}rgen Randrup,
Hideo Suganuma, 
Yoshitaka Hatta
and
Yuto Mori
for useful discussions.
This work is supported in part by the Grants-in-Aid for Scientific Research
 from JSPS (Nos. 
19K03872, 
19H01898 
and
19H05151) 
and by the Yukawa International Program for Quark-hadron Sciences (YIPQS).

\appendix
\section{Matsubara frequency summation}
\label{Sec:Matsubara}

In deriving Eq.~\eqref{Eq:Zfree}, we have used the Matsubara frequency
summation formulae,
\begin{align}
S=T\sum_n g(\omega_n=2\pi nT)
=-i \sum_{\omega_0} \frac{\mathrm{Res}\, g(\omega_0)}{e^{i\omega_0/T}-1}
\ ,
\label{Eq:Matsubara}
\end{align}
where $g(\omega)$ is an analytic function of $\omega$,
does not have poles on the real axis,
and decreases faster than $1/\omega$ at $\omega \to \infty$,
{\em i.e.}
$\lim_{|\omega|\to \infty} \omega g(\omega) =0$.
The poles and residues of $g(\omega)$ are denoted by
$\omega_0$ and $\mathrm{Res}\,g(\omega_0)$.

Specifically, we consider the following sum
\begin{align}
S=\frac{1}{N}\sum_n\log\left[E^2+\sin^2(\omega_n/2)\right]
\ ,
\label{Eq:MatsubaraS}
\end{align}
where $T=1/N$ and $\omega_n=2\pi n/N$.
The derivative $dS/dE$ is in the form of Eq.~\eqref{Eq:Matsubara},
\begin{align}
\frac{dS}{dE}
=&\frac{1}{N}\sum_n \frac{2E}{E^2+\sin^2(\omega_n/2)}
\ ,
\end{align}
where $g(\omega)=2E/(E^2+4\sin^2(\omega/2))$ and $T=1/N$.
The poles and residues of $g(\omega)$ are found to be
$i\omega_0=\pm \Omega = \pm 2\mathrm{arcsinh}\,E$
and
$\mathrm{Res}\,g(\omega_0)=\pm 2i/\sqrt{1+E^2}$,
so $dS/dE$ is obtained as
\begin{align}
\frac{dS}{dE}
=& \frac{2}{\sqrt{1+E^2}}\,\frac{e^{N\Omega}+1}{e^{N\Omega}-1}
= \frac{2\coth(N\Omega/2)}{\sqrt{1+E^2}}
\ .
\end{align}
By integration, the sum in Eq.~\eqref{Eq:MatsubaraS} is found to be
\begin{align}
S
=& \frac{2}{N}\log\left[\sinh(N\Omega/2)\right]+\mathrm{const.}
\ .
\end{align}
The constant can be fixed as $2\log2(1/N-1)$
by considering the large $E$ limit of $S$, 
$S = 2\log{E}+\mathcal{O}(1/E)$.
By substituting $E=\omega_{\bm{k}}/2\xi$, we obtain Eq.~\eqref{Eq:Zfree}.

The same formula can be used to obtain $\Delta=\VEVev{\phi^2}$ 
in Eq.~\eqref{Eq:Delta}.
\begin{align}
\Delta_{\bm{k}} 
=& \frac{1}{4\xi}
\frac{1}{N}
\sum_{n}
\frac{1}{(\omega_{\bm{k}}/2\xi)^2+\sin^2(\omega_n/2)}
=\frac{1}{4\xi}
\frac{\coth(\Omega_{\bm{k}}/2T)}
{(\omega_{\bm{k}}/2\xi)\sqrt{1+(\omega_{\bm{k}}/2\xi)^2}}
\nonumber\\
=&\frac{1}{L^3}
\frac{1}{\omega_{\bm{k}}\sqrt{1+(\omega_{\bm{k}}/2\xi)^2}}
\left[
\frac12+\frac{e^{-\Omega_{\bm{k}}/T}}{1-e^{-\Omega_{\bm{k}}/T}}
\right]
\ ,
\end{align}
where $\Omega_{\bm{k}}=2\xi\mathrm{arcsinh}(\omega_{\bm{k}}/2\xi)$.
Then we can obtain $\Delta=\VEVev{\phi^2}=\sum_{\bm{k}}\Delta_{\bm{k}}/L^3$
in Eq.~\eqref{Eq:Delta}.

\section{Fourier transform of time-correlation function}
\label{Sec:spec}

\begin{figure}[htbp]
\begin{center}
\includegraphics[width=7.5cm,bb=0 0 360 250]{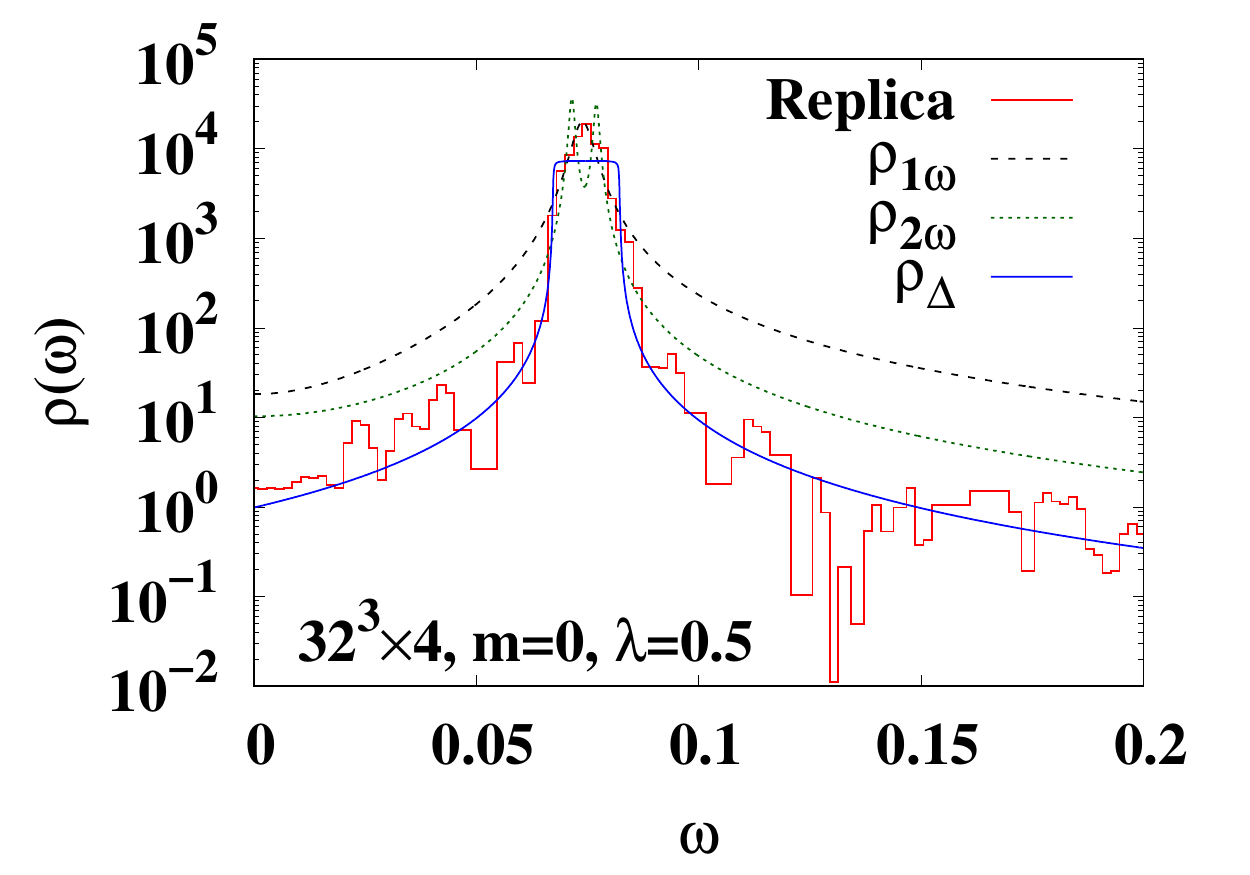}%
\includegraphics[width=7.5cm,bb=0 0 360 250]{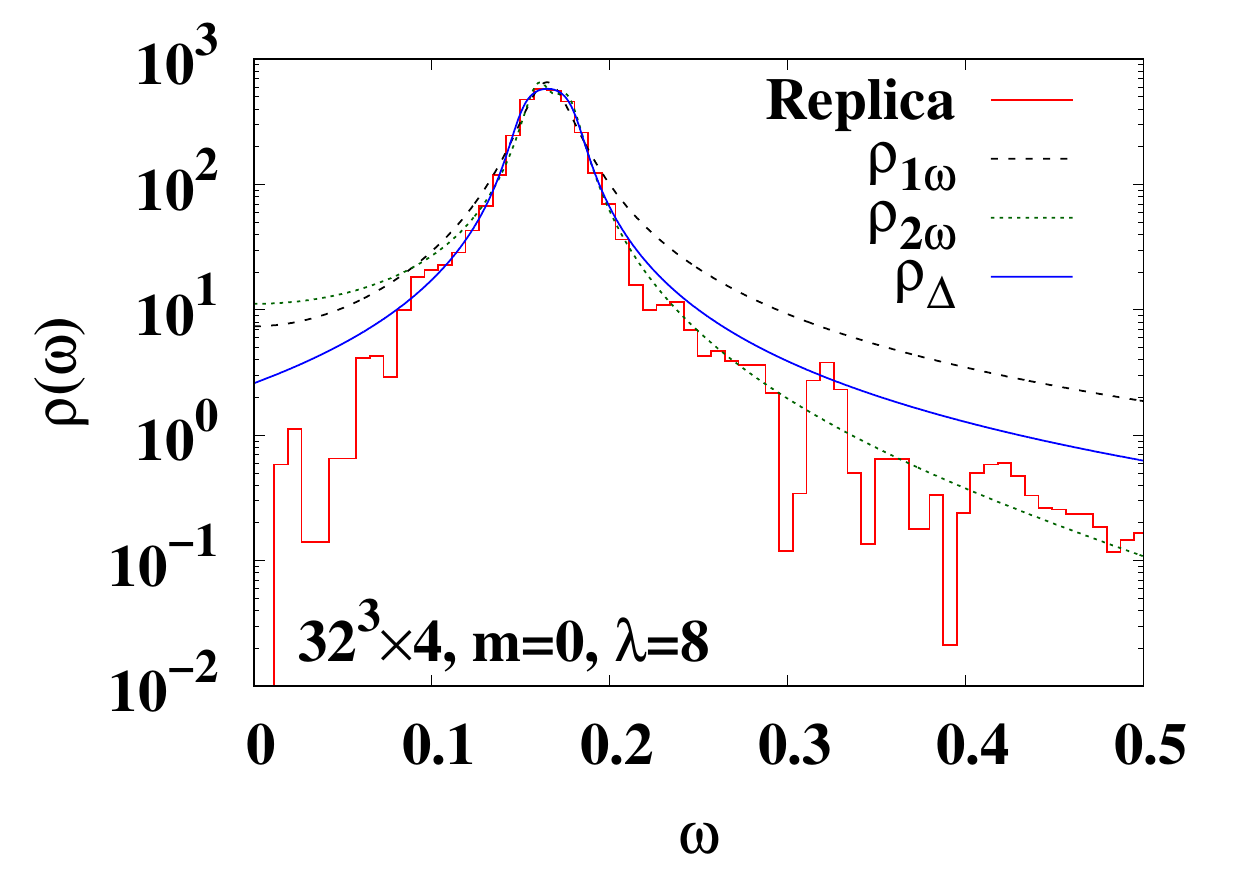}
\end{center}
\caption{Fourier transform of the time-correlation function.
We show the results with mass counterterm at $\lambda=0.5$ (left)
and $8$ (right) with $m=0$.
Histograms show the Fourier transform of $C(t)$
obtained from the replica evolution.
Solid, dashed and dotted curves show the Fourier transform $\rho_\Delta(\omega)$
and the Fourier transform of $f_{1\omega}$ and $f_{2\omega}$, respectively.
}\label{Fig:spec}
\end{figure}

In Sec. \ref{Sec:Results}, the damping rate is found to be larger
in the weak coupling region with mass counterterm at $m=0$.
Here we would like to discuss this point
by using the Fourier transform of the time-correlation function,
\begin{align}
\rho(\omega) \equiv \frac12\int_{-\infty}^\infty dt e^{i\omega t}C(|t|) .
\end{align}
In Fig. \ref{Fig:spec}, we show $\rho(\omega)$
obtained from the replica evolution with mass counterterm at $\lambda=0.5$
and $8$ with $m=0$
in comparison with the Fourier transform of
the fitting function $f_{1\omega}$.
At $\lambda=0.5$, the spectrum has a peak having the width 
of the order of $10^{-2}$, but the tails fall off much faster
than the behavior expected from the width or the damping rate
from the fitting function 
$f_{1\omega}$, $\gamma\simeq 2.6\times 10^{-3}$.

One of the possible interpretations of this spectrum
is to consider that there are several modes,
each of which has a small damping rate
but has a mass spread in the region with $10^{-2}$ width.
As an attempt, we use the two damped oscillator functions,
where the two frequencies are close to each other,
\begin{align}
f_{2\omega}(t)
=&A \exp(-\gamma t)
\left\{
r \cos[(\omega+\delta\omega)t+\delta]
+(1-r) \cos[(\omega-\delta\omega)t+\delta]
\right\}
\ .
\label{Eq:f2w}
\end{align}
The dotted lines show the Fourier transform of
$f_{2\omega}$ fitted to $C(t)$.
The tail region is found to be suppressed, but not enough at $\lambda=0.5$.
Next, we consider the following fitting function,
convolution of the step function and the Lorentzian, for the Fourier transform,
\begin{align}
\rho_{\Delta}(\omega)=&\frac{A}{2\Delta}\int_{M-\Delta}^{M+\Delta}\,
\frac{\gamma dM'}{(\omega-M')^2+\gamma^2}
\nonumber\\
=&\frac{A}{2\Delta}\left[
\arctan\left(\frac{\omega-M+\Delta}{\gamma}\right) 
-\arctan\left(\frac{\omega-M-\Delta}{\gamma}\right) 
\right] .
\end{align}
The fitting results are shown by the solid curves,
which shows both wide width of the peak and the fast fall off.
The damping rate of each mode becomes smaller,
$\gamma\simeq 1.5\times 10^{-4}$, 
and roughly agrees with the perturbative estimate,
as shown by open triangles in Fig. \ref{Fig:gam}.

At larger coupling, fitting results of the damping rate
with $f_{1\omega}$, $f_{2\omega}$ and $\rho_{\Delta}$ are consistent,
and the peak part of the spectrum is reproduced in these functions,
as shown in the right panel of Fig. \ref{Fig:spec}.

\section{Time-odd part of the time-correlation function in harmonic oscillator}
\label{Sec:PB}
While the expectation value of symmetrized (Weyl ordered) product
can be obtained in classical dynamics,
we need additional care to evaluate the expectation value 
of the anti-symmetrized product such as the commutator.
For example, the time-odd part of the time-correlation function
may be obtained by using the quantum-classical correspondence 
for the commutator~\cite{Aarts-spec},
\begin{align}
[A,B] \to i\hbar\{A,B\}_\mathrm{PB} + \mathcal{O}(\hbar^3)
\ ,
\end{align}
where we explicitly show $\hbar$ here and $\{A,B\}_\mathrm{PB}$
is the Poisson bracket.
If we ignore $\mathcal{O}(\hbar^3)$,
the time-odd part of the time-correlation function would be obtained as
\begin{align}
\VEVT{\frac12[\hat{x}_H(t),\hat{x}_H(0)]}
\simeq& \VEV{\frac{i}{2}\{x(t),x(0)\}_\mathrm{PB}}
= \frac{i}{2}\VEV{
\sum_{\tau,\tau'}
\left[
 \frac{\partial x_\tau(t)}{\partial x_{\tau'}(t_0)}
 \frac{\partial x_\tau(0)}{\partial p_{\tau'}(t_0)}
-\frac{\partial x_\tau(t)}{\partial p_{\tau'}(t_0)}
 \frac{\partial x_\tau(0)}{\partial x_{\tau'}(t_0)}
\right]}
\nonumber\\
=& \frac{i}{2}\VEV{
\sum_{n,n'}
\left[
 \frac{\partial \bar{x}_n(t)}{\partial \bar{x}_{n'}(t_0)}
 \frac{\partial \bar{x}_n(0)}{\partial \bar{p}_{n'}(t_0)}
-\frac{\partial \bar{x}_n(t)}{\partial \bar{p}_{n'}(t_0)}
 \frac{\partial \bar{x}_n(0)}{\partial \bar{x}_{n'}(t_0)}
\right]}
\nonumber\\
=&-\frac{i}{2}\sum_n \frac{1}{M_n}\sin M_nt
\label{Eq:PB}
\ .
\end{align}
Since the Fourier transformation and the time evolution
are the canonical transformation, we can choose 
either $(x_\tau,p_\tau)$ or $(\bar{x}_n,\bar{p}_n)$
in calculating the Poisson bracket
and the time $t_0$ should be arbitrary.
The zero Matsubara frequency contribution in Eq.~\eqref{Eq:PB}
agrees with the quantum mechanical result.

In the case of coupled oscillators such as the field theories,
it is in principle possible to calculate the Poisson bracket
of the unequal-time observables
by using the Hessian matrix~\cite{Entropy},
while it requires to store the matrix elements of degrees of freedom squared,
$(2N_\mathrm{dof})^2$ with $N_\mathrm{dof}=NL^3$ for one component
scalar field theory on the $L^3\times N$ lattice.
We also need to multiply the matrix at each step of time,
so the numerical cost is much larger than the time-correlation function
discussed in this article.



\begin{thebibliography}{99}
\bibitem{TDGP}
E. P. Gross, Il Nuovo Cimento {\bf 20}, 454 (1961);
L. P. Pitaevskii, Sov. Phys. JETP {\bf 13}, 451 (1961);
P. Muruganandam, S. K. Adhikari,
Comp. Phys. Comm. {\bf 180}, 1888 (2009).

\bibitem{TDHF}
D. J. Thouless and J. G. Valatin,
Nucl. Phys. {\bf 31} 211 (1962);
%
A. D. McLachlan and M. A. Ball, Rev. Mod. Phys. {\bf 36}, 844 (1964);
Y. M. Engel, D. M. Brink, K. Goeke, S. J. Krieger, D. Vautherin,
Nucl. Phys. A {\bf 249}, 215 (1975).


\bibitem{inflation}
K. Sato, Mon. Notices Royal Astron. Soc. {\bf 195}, 467 (1981);
A. H. Guth, Phys. Rev. D {\bf 23}, 347 (1981).

\bibitem{CSS}
  S. Y. Khlebnikov and I. I. Tkachev,
  Phys. Rev. Lett.  {\bf 77}, 219 (1996).

\bibitem{CYM}
L. D. McLerran and R. Venugopalan,
Phys. Rev. D {\bf 49}, 2233 (1994).
  P. Romatschke and R. Venugopalan,
  Phys. Rev. Lett.  {\bf 96}, 062302 (2006);
  T. Lappi and L. McLerran,
  Nucl. Phys. A {\bf 772}, 200 (2006);
  J. Berges, S. Scheffler and D. Sexty,
  Phys. Rev. D {\bf 77}, 034504 (2008);
  K. Fukushima and F. Gelis,
  Nucl. Phys. A {\bf 874}, 108 (2012).
T. Epelbaum and F. Gelis,
Phys. Rev. Lett. {\bf 111}, 232301 (2013).

\bibitem{Dumitru-Nara}
  A. Dumitru and Y. Nara,
  Phys. Lett. B {\bf 621}, 89 (2005);
  A. Dumitru, Y. Nara and M. Strickland,
  Phys. Rev. D {\bf 75}, 025016 (2007).

\bibitem{YM-chaos}
  S. G. Matinyan, E. B. Prokhorenko and G. K. Savvidy,
  JETP Lett.  {\bf 44}, 138 (1986);
%
B. M\"uller and A. Trayanov,
Phys. Rev. Lett. {\bf 68}, 3387 (1992).
%

\bibitem{Entropy}
T. Kunihiro, B. Muller, A. Ohnishi, A. Schafer, T. T. Takahashi and A. Yamamoto,
Phys. Rev. D {\bf 82}, 114015 (2010); 
H. Iida, T. Kunihiro, B. Mueller, A. Ohnishi, A. Schaefer and T. T. Takahashi,
Phys. Rev. D {\bf 88}, 094006 (2013); 
H. Tsukiji, H. Iida, T. Kunihiro, A. Ohnishi and T. T. Takahashi,
Phys. Rev. D {\bf 94}, 091502 (2016); 
H. Tsukiji, T. Kunihiro, A. Ohnishi and T. T. Takahashi,
PTEP {\bf 2018}, 013D02 (2018). 

\bibitem{Shear}
M. M. Homor and A. Jakovac,
Phys. Rev. D {\bf 92}, 105011 (2015);
%

\bibitem{Matsuda}
H. Matsuda, T. Kunihiro, A. Ohnishi, T. T. Takahashi,
PTEP \textbf{2020}, 053D03 (2020);
%
H. Matsuda, T. Kunihiro, A. Ohnishi, T. T. Takahashi,
arXiv:2007.06886 [hep-ph].


\bibitem{Vlasov}
A. A. Vlasov, J. Exp. Theor. Phys. {\bf 8}, 291 (1938);
Soviet Physics Uspekhi, {\bf 10}, 721 (1968).

\bibitem{vonNeumann}
John von Neumann, G{\"o}ttinger Nachrichten {\bf 1}, 245 (1927).

\bibitem{Wigner}
E. Wigner, Phys. Rev. {\bf 40}, 749 (1932).

\bibitem{TP} 
N. Rostoker and M. N. Rosenbluth, Phys. Fluids {\bf 3}, 1 (1960);\\
  C. Y. Wong,
  Phys. Rev. C {\bf 25}, 1460 (1982).

\bibitem{Bodeker:1995pp} 
  D. Bodeker, L. D. McLerran and A. V. Smilga,
  Phys. Rev. D {\bf 52}, 4675 (1995).

\bibitem{Greiner:1996dx} 
  C. Greiner and B. Muller,
  Phys. Rev. D {\bf 55}, 1026 (1997).

\bibitem{Aarts}
  G. Aarts and J. Smit,
  Phys. Lett. B {\bf 393}, 395 (1997);
  Nucl. Phys. B {\bf 511}, 451 (1998);
  G. Aarts, G. F. Bonini and C. Wetterich,
  Phys. Rev. D {\bf 63}, 025012 (2001).

\bibitem{Berges:2004yj} 
  J. Berges,
  AIP Conf. Proc.  {\bf 739}, 3 (2004).

\bibitem{Aarts:2001yn} 
  G. Aarts and J. Berges,
  Phys. Rev. Lett.  {\bf 88}, 041603 (2002).

\bibitem{Hatta-Nishiyama} 
  Y. Hatta and A. Nishiyama,
  Nucl. Phys. A {\bf 873}, 47 (2012).
\bibitem{QStat}
E. A. Uehling and G. E. Uhlenbeck, Phys. Rev. {\bf 43}, 552 (1933);
%
G. F. Bertsch and S. Das Gupta,
Phys. Rept. \textbf{160}, 189 (1988);
%
  A. Ono, H. Horiuchi, T. Maruyama and A. Ohnishi,
  Phys. Rev. Lett.  {\bf 68}, 2898 (1992);
%
  Prog. Theor. Phys.  {\bf 87}, 1185 (1992);
%
%
  A. Ohnishi and J. Randrup,
  Phys. Rev. Lett.  {\bf 75}, 596 (1995);
%
%
  Phys. Lett. B {\bf 394}, 260 (1997);
%
A. Ono and H. Horiuchi,
Phys. Rev. C \textbf{53}, 2958 (1996);
%
  Y. Hirata, Y. Nara, A. Ohnishi, T. Harada and J. Randrup,
  Prog. Theor. Phys.  {\bf 102}, 89 (1999);
%
P. Chomaz, M. Colonna and J. Randrup,
Phys. Rept. \textbf{389}, 263 (2004).
\bibitem{SQ}
  G. Parisi and Y. s. Wu,
  Sci. Sin.  {\bf 24}, 483 (1981);
  G. Parisi,
  Phys. Lett.  {\bf 131B}, 393 (1983);\\
  P. H. Damgaard and H. Huffel,
  Phys. Rept.  {\bf 152}, 227 (1987).

\bibitem{HMC}
  S. Duane, A. D. Kennedy, B. J. Pendleton and D. Roweth,
  Phys. Lett. B {\bf 195}, 216 (1987).

\bibitem{Kapusta}
  J. I. Kapusta and C. Gale,
  "Finite-temperature field theory: Principles and applications"
  (Cambridge University Press, Cambridge, 2006).

\bibitem{TwoLoop} 
  R. R. Parwani,
  Phys. Rev. D {\bf 45}, 4695 (1992)
  Erratum: [Phys. Rev. D {\bf 48}, 5965 (1993)].

\bibitem{Poincare}
H. Poincar{\'e},
Acta Math. {\bf 13}, 1 (1890).

\bibitem{Aarts-spec}
  G. Aarts,
  Phys. Lett. B {\bf 518}, 315 (2001).

\bibitem{Mueller:2002gd}
A.~H.~Mueller and D.~T.~Son,
Phys. Lett. B \textbf{582}, 279 (2004).

\bibitem{NSO}
  D. N. Zubarev, A. V. Prozorkevich, and S. A. Smolyanskii,
  Theor. Math. Phys. {\bf 40}, 821 (1979);
%
  F. Becattini, L. Bucciantini, E. Grossi, and L. Tinti,
  Eur. Phys. J. C {\bf 75}, 191 (2015);
%
  S.-i. Sasa,
  Phys. Rev. Lett. {\bf 112}, 100602 (2014);
  T. Hayata, Y. Hidaka, T. Noumi and M. Hongo,
  Phys. Rev. D {\bf 92}, 065008 (2015).

\end{thebibliography}
\end{document}